\documentstyle[preprint,aps]{revtex}

\begin{document}

\draft

\preprint{January, 2001}

\title{Yang-Mills Instantons Sitting on a Ricci-flat Worldspace \\ 
       of Double D4-brane}

\author{Hongsu Kim\footnote{e-mail : hongsu@hepth.hanyang.ac.kr} and
Yongsung Yoon\footnote{e-mail : cem@hepth.hanyang.ac.kr}}

\address{Department of Physics \\
Hanyang University, Seoul, 133-791, KOREA}


\maketitle

\begin{abstract}
Thus far, there seem to be no complete criteria that can settle the issue as to what 
the correct generalization of the Dirac-Born-Infeld (DBI) action, describing the low-energy 
dynamics of the $D$-branes, to the non-abelian case would be. According to recent suggestions, one
might pass the issue of worldvolume solitons from abelian to non-abelian setting by considering 
the stack of multiple, coincident $D$-branes and use it as a guideline to construct or censor 
the relevant non-abelian version of the DBI action. In this spirit, here we are interested
in the explicit construction of $SU(2)$ Yang-Mills (YM) instanton solutions in the background 
geometry of two coincident probe $D4$-brane worldspaces particularly when
the metric of target spacetime in which the probe branes are embedded is given by the
Ricci-flat, magnetic extremal 4-brane solution in type IIA supergravity theory with its worldspace 
metric being given by that of Taub-NUT and  Eguchi-Hanson solutions, the two best-known gravitational 
instantons. And then we demonstrate that with this YM instanton-gravitational instanton configuration
on the probe $D4$-brane worldvolume, the energy of the probe branes attains its minimum value and hence
enjoys stable state provided one employs the Tseytlin's non-abelian DBI action for the description 
of multiple probe $D$-branes. In this way, we support the arguments in the literature in favor of 
Tseytlin's proposal for the non-abelian DBI action.
\end{abstract}


\narrowtext

\begin{center}
{\rm \bf I. Introduction}
\end{center}

With the recent advent of the physics of $D$-branes [1], the solitons in the non-perturbative
spectrum of string theory, it has been realized that their low-energy dynamics can
be properly described by the so-called, ``Dirac-Born-Infeld (DBI)'' action [1,2].
Since single branes are known to be described by the abelian DBI action, one might
naturally expect that the multiple branes [3] would be by some non-abelian generalization
of the DBI action. Unfortunately, however, there seem to be no complete criteria
that can settle the issue as to what the correct generalization of the
DBI action to the non-abelian case would be. 
To pass from the abelian version to the non-abelian one, the first 
obvious treatment to be done is to render the action group scalar by taking the
trace over the group indices appearing in the non-abelian gauge field strength term.
Indeed, different group trace operations have been proposed in the literature and
they include ; $Tr$, the ordinary trace [1], $Str$, the symmetrized trace [4] and
$(Str + i~Atr)$, the combination of the symmetrized and the antisymmetrized trace [5].
Here, we will argue that the symmetrized trace operation put forward by Tseytlin
stands out particularly in association with the ``BPS'' solitons that the $D$-branes
may admit on their worldvolumes. Actually, it shall be demonstrated that Tseytlin's
choice of non-abelian DBI action, but no others,  ``knows'' about the energy-minimizing 
worldvolume solitons. Since we shall take the BPS soliton on the $D$-brane worldvolume 
as a mean to argue the relevance of Tseytlin's version of non-abelian DBI action, it seems
appropriate to remind briefly the status of worldvolume solitons uncovered in the literature
thus far. It has been found that the $D4$-brane worldvolume theories admit abelian
instantons [6] ; those of $D3$-brane admit abelian monopoles and dyons [7] ; those of
$D2$-brane admit abelian vortices [7,8] ; and lastly, those of $D$-string admit kinks [9].
Then taking the abelian DBI action for the worldvolume field theory, it has been realized 
that the BPS-type condition for each case minimizes the energy of the corresponding
worldvolume soliton. And another interesting point is that all of the above configurations
can be obtained by starting from the $D4$-brane case and successively applying $T$-duality [10].
Thus it seems quite natural to pass this issue of worldvolume solitons from abelian to
non-abelian setting by considering instead the stack of multiple, coincident $D$-branes
and use it as a guideline to construct or censor the non-abelian version of the DBI action
and this is precisely what we would like to do in the present work.
Then what is so special about Tseytlin's suggestion for the non-abelian DBI action ?
As we shall see in a moment, for non-abelian gauge field configuration satisfying the (anti)
self-dual field strength $F_{\alpha\beta} = \pm \tilde{F}_{\alpha\beta}$ (with $\alpha$,
$\beta$ being the $D4$-brane {\it worldspace} indices), the determinant in the DBI action can
be written as a ``complete square'' {\it linearizing} the DBI action and then turning it into 
that of usual Yang-Mills theory [10,11]. What is more, since we are dealing with static soliton 
configurations, the energy density of the $D4$-brane is just $H_{DBI} = - L_{DBI}$, which gets 
{\it minimized} if and only if $F_{\alpha\beta} = \pm \tilde{F}_{\alpha\beta}$ [6,10,11]. 
Thus the BPS condition,  
or the (anti) self-duality of non-abelian gauge theory instanton solution at once linearizes
the otherwise highly non-linear DBI action and minimizes the energy of the $D4$-brane provided
Tseytlin's non-abelian DBI action is employed. In order to demonstrate that this procedure
can indeed work, the question would be whether or not one can actually construct, say, 
$SU(2)$ Yang-Mills instanton solution having (anti) self-dual field strength in the
background of $D4$-brane worldspace. Thus at this point, it might be relevant to remind what 
has been done and suggested to be done in the literature to answer to this question. 
In [10], it has been pointed out that a stack of two coincident $D4$-branes 
with {\it flat} worldvolume geometry admits worldvolume solitons corresponding to the standard 
BPST [12] $SU(2)$ 
YM instanton solutions to the (anti) self-dual YM equation in the {\it flat} $D4$-brane worldspace.
And then in [11], it was just {\it argued} that in a similar manner, a pile of two coincident 
$D4$-branes with {\it Ricci-flat} worldvolume geometry may admit worldvolume solitons this time 
corresponding to the $SU(2)$ YM instanton solutions to the (anti) self-dual YM equation in the 
background of {\it Ricci-flat} $D4$-brane worldspace described, say, by 4-dimensional gravitational
instantons. In [11], however, they were not able to explicitly demonstrate the construction of such
YM instanton solutions.
Thus here in this work, we would like to demonstrate in an explicit manner that
the worldvolume solitons of this sort, i.e., $SU(2)$ YM instanton solutions in the background
of {\it Ricci-flat} $D4$-brane worldspace with its metric being described by the Taub-NUT or
Eguchi-Hanson metric can actually be constructed. To be a little more precise, we shall be interested 
in the explicit construction of $SU(2)$ Yang-Mills instanton solutions in the background geometry of a
stack of the two coincident probe $D4$-brane worldspaces particularly when
the metric of target spacetime in which the probe branes are embedded is given by the
Ricci-flat, magnetic extremal 4-brane solution in type IIA theory with its worldspace metric
being given by that of Taub-NUT or Eguchi-Hanson solution, the two best-known
gravitational instantons. And then we shall
demonstrate that with this YM instanton-gravitational instanton configuration on the probe
$D4$-brane worldvolume, the energy of the probe branes attains its minimum value and hence
enjoys stable state provided one employs the Tseytlin's non-abelian DBI
action for the description of multiple probe $D$-branes.
On the technical side, then, it might be relevant to describe how to generally construct 
the $SU(2)$ YM instantons as solutions to (anti) self-dual YM equations in the background of 
typical 4-dimensional gravitational instantons. Generally speaking, well below the Planck scale, 
the strength of gravity is negligibly small relative to those of particle physics
interactions described by non-abelian gauge theories. Thus one might overlook the effects of gravity 
at the elementary particle physics scale. Nevertheless, as far as the topological aspect is concerned,
gravity may have marked effects even at the level of elementary particle physics. 
Namely, the non-trivial topology of the gravitational field may play a role crucial enough to 
dictate the topological properties of, say, $SU(2)$ Yang-Mills (YM) gauge field [12] as has been 
pointed out long ago [13]. Being an issue of great physical interest and importance, quite a few 
serious study along this line have appeared in the literature but they were restricted to the background
gravitational field with high degree of isometry such as the Euclideanized Schwarzschild geometry [13] or
the Euclidean de Sitter space [14]. Even the works involving more general background spacetimes 
including gravitational instantons (GI) were mainly confined to the case of asymptotically-
locally-Euclidean (ALE) spaces which is one particular such GI and employed rather indirect and
mathmatically-oriented solution generating methods such as the ADHM construction [15]. 
Recently, we [16] have proposed a ``simply physical'' and perhaps the most direct algorithm for 
generating the YM instanton solutions in all species of known GI. Thus in the present work, 
we would like to employ this recently developed method to construct $SU(2)$ YM instanton solutions 
in the background of {\it Ricci-flat} $D4$-brane worldspace described by 4-dimensional 
gravitational instantons such as Taub-NUT (TN) or Eguchi-Hanson (EH) metric. \\
The rest of the paper is organized as follows. In sect.II, we shall present the Ricci-flat extremal
$p$-brane solutions in supergravity (SUGRA) theories. Although their existence has been pointed
out and the solution forms written down in [11], a careful and explicit derivation is hardly available
in the literature.\footnote{A rather concise description of the derivation has been given by 
B. Janssen (JHEP {\bf 0001}, 044 (2000)) and we were informed (private communication) that also
it can be found in the PhD thesis of D. Brecher although the latter is not generally available.}
Thus we provide the detailed derivation in this section as it plays the central role in this work.
Sect.III is devoted to the introduction of different versions of the non-abelian DBI action and 
the detailed description of the features of Tseytlin's action particularly in connection to the
worldspace instanton of the $D4$-brane. Sect.IV is the main part of this work containing new
ingredients. Namely, there we discuss the construction of $SU(2)$ instanton solutions in the
background of TN and EH metrics representing the Ricci-flat $D4$-brane worldspace. Lastly in
sect.V, we conclude with discussions and in the Appendix, we provide detailed
analysis of the interesting nature of these YM instantons obtained in sect.IV.

\begin{center}
{\rm \bf II. Ricci-flat extremal $p$-brane solutions in supergravity theories}
\end{center}

{\rm \bf 1. Construction of solutions}

In order to derive and study the $M$-brane solutions in (the bosonic sector of)
$D=11$ SUGRA and $D$-brane solutions in $D=10$ type IIA/IIB SUGRA conveniently
in a single setting, we would like to consider a system in generally $D$-spacetime
dimensions comprising the metric $G_{MN}$, a scalar (dilaton) field $\Phi$, and an $(n-1)$-form
antisymmetric R-R tensor gauge field $A_{[n-1]}$ with the associated field strength
$F_{[n]}$ described by the action (our method here to construct the solutions closely follows
that in [17])
\begin{eqnarray}
S = \int d^{D}X \sqrt{G} [R - {1\over 2}\nabla_{M}\Phi \nabla^{M}\Phi - {1\over 2n!}
e^{\tilde{a}\phi}F^2_{[n]}].
\end{eqnarray}
Upon extremizing this action with respect to $G_{MN}$, $\Phi$ and $A_{M_{1}...M_{n-1}}$,
one would get a set of classical field equations of which the concrete forms shall be
given later on. (Of course, it is known that the field equations of type IIB theory cannot 
be derived from a covariant action. Nevertheless, the field equations that result from
the action given above are general ones in that they involve those of type IIB theory.)
In order to practically solve these equations of motion, we would have
to make a simplifying ans\H{a}tz for solutions. As a simplest choice, one normally require
translational symmetry along the $(p+1)=(n-1)\equiv d$-dimensional worldvolume of the
$p$-brane configuration and isotropy in the directions ``transverse'' to this $p$-brane
worldvolume, namely $(Poincare)_{d}\times SO(D-d)$-symmetric solution ans\H{a}tz. 
Accordingly, then, the spacetime coordinates can be split into two ranges
\begin{eqnarray}
X^{M} = (x^{\mu}, ~y^{m})
\end{eqnarray}
where $x^{\mu}$$(\mu = 0,1,...,p=(d-1))$ are the coordinates on the $p$-brane worldvolume
and $y^{m}$$(m = (p+1),...,(D-1))$ are the coordinates transverse to the 
worldvolume. Thus the worldvolume has $(p+1)=d$ dimensions and the number of transverse
directions is $(D-d)\equiv (\tilde{d}+2)$.
Of course this highly symmetric restriction on the solution ans\H{a}tz can be
somewhat relaxed in more generalized versions of the class of $p$-brane solutions
and particularly here in the present work, we start by abandoning the Poincare-symmetry
on the brane worldvolume and end by demonstrating that a simple, ``Ricci-flat''
$p$-brane solutions can actually be constructed. Namely, we employ the ans\H{a}tz for the
spacetime metric as
\begin{eqnarray}
ds^2 &=& G_{MN}dX^{M}dX^{N} \\
&=& e^{2A(r)}\gamma_{\mu\nu}(x)dx^{\mu}dx^{\nu} + e^{2B(r)}\delta_{mn}
dy^{m}dy^{n} \nonumber \\
&=& e^{2A(r)}\hat{\eta}_{ab}\hat{e}^{a}_{\mu}(x)\hat{e}^{b}_{\nu}(x)dx^{\mu}dx^{\nu} + e^{2B(r)}
dy^{m}dy^{m} \nonumber \\
&=& \eta_{IJ}e^{I}(X)e^{J}(X) \nonumber
\end{eqnarray}
where $r=(y^{m}y^{m})^{1/2}$ is the isotropic radial coordinate in the space
transverse to the brane worldvolume and we introduced, to carry out later on the computation 
of the Ricci tensor components $R_{MN}$ in a easier and elegant manner, the orthonormal basis or
the vielbein $e^{I}(X)=e^{I}_{M}dX^{M}=e^{I}_{\mu}dx^{\mu}+e^{I}_{m}dy^{m}$
given by
\begin{eqnarray}
e^{I}(X) = \{e^{a}=e^{A(r)}\hat{e}^{a}(x)=e^{A(r)}\hat{e}^{a}_{\mu}(x)dx^{\mu},
~~e^{m}=e^{B(r)}dy^{m}\}                                                                                 
\end{eqnarray}
and to summarize, here $M=(\mu, ~m)$ are coordinate basis indices and $I=(a, ~m)$ are
non-coordinate, orthonormal frame indices with $\mu, ~a = 0,1,...,p=(d-1)$ and
$m = (p+1),...,(D-1)$ respectively. Since the metric components depend only on ``$r$'',
the $SO(D-d)$-symmetry in the transverse directions still remains. Then the corresponding
ans\H{a}tz for the scalar dilaton field is simply
\begin{eqnarray}
\Phi = \Phi (r).
\end{eqnarray}
Finally, for the $(p+1)=(n-1)$-form antisymmetric R-R tensor gauge field  
$A_{M_{1}...M_{n-1}}$, two kinds of ansatz related by duality transformation are possible.
Obviously, the first possibility is to choose $A_{[n-1]}$ such that it ``supports'' a
$(p+1)=(n-1)$-dimensional worldvolume. To be a little more concrete, we naturally expect that
$A_{[n-1]}$ couples directly to the $(p+1)=(n-1)$-dimensional worldvolume of the $p$-brane,
just as the 1-form Maxwell gauge potential couples to the worldline of a charged particle
as $eA_{\mu}dx^{\mu}$ with $e$ being the gauge charge. Then the ``charge'' of the $p$-brane
here will be obtained via Gauss' law from the surface integral involving the associated
field strength $F_{[n]}$. This first possible choice shall be referred to as ``electric
(or elementary)'' ans\H{a}tz and amounts to
\begin{eqnarray}
A_{\mu_{1}...\mu_{n-1}} = \epsilon_{\mu_{1}...\mu_{n-1}}e^{C(r)}
\end{eqnarray}
with others being zero or in terms of its field strength
\begin{eqnarray}
F_{m\mu_{1}...\mu_{n-1}} = \epsilon_{\mu_{1}...\mu_{n-1}}\partial_{m}e^{C(r)}
\end{eqnarray}
with all others zero. Here, $\epsilon_{\mu_{1}...\mu_{n-1}}$ is generally the curved 
spacetime version of the totally antisymmetric tensor.
We now turn to the second possibility for the choice of the ans\H{a}tz
for $A_{[n-1]}$. This second possibility can be most conveniently expressed in terms of
the field strength as it can be obtained by considering the Hodge dualized field strength
$\tilde{F}_{[n]}$, which is a $(D-n)=(D-d-1)=(D-p-2)$-form. Since the duality transformation
is involved, this second choice may be called ``magnetic (or solitonic)'' ans\H{a}tz and
amounts to 
\begin{eqnarray}
F_{m_{1}m_{2}...m_{(D-n)}} = \lambda \epsilon_{m_{1}m_{2}...m_{(D-n)}l}
{y^{l}\over r^{(D-n+1)}}
\end{eqnarray}
where the undetermined parameter $\lambda$ in this ansatz is an integration constant
representing the magnetic charge. 
Having constructed two kinds of ans\H{a}tz for the R-R tensor gauge field, we now can write
down the classical field equations in an explicit manner
\begin{eqnarray}
R_{MN} &=& {1\over 2}\partial_{M}\Phi \partial_{N}\Phi + S_{MN} ~~~{\rm with} \nonumber \\
S_{MN} &=& {1\over 2(n-1)!}e^{\tilde{a}\Phi}\left[F_{M...}F_{N}^{...} - {(n-1)\over n(D-2)}
G_{MN}F^2\right], \nonumber \\
{1\over \sqrt{G}}&&\partial_{M}[\sqrt{G}G^{MN}\partial_{N}\Phi] = {\tilde{a}\over 2n!}
e^{\tilde{a}\Phi}F^2, \nonumber \\
{1\over \sqrt{G}}&&\partial_{M_{1}}[\sqrt{G}e^{\tilde{a}\Phi}F^{M_{1}...M_{n}}] = 0 
\end{eqnarray}
for the electric ans\H{a}tz and with
\begin{eqnarray}
S_{MN} &=& {1\over 2(D-n-1)!}e^{\tilde{a}\Phi}\left[F_{M...}F_{N}^{...} - {(D-n-1)\over (D-n)(D-2)} 
G_{MN}F^2\right], \nonumber \\
{1\over \sqrt{G}}&&\partial_{M}[\sqrt{G}G^{MN}\partial_{N}\Phi] = {\tilde{a}\over 2(D-n)!}
e^{\tilde{a}\Phi}F^2, \nonumber \\
{1\over \sqrt{G}}&&\partial_{M_{1}}[\sqrt{G}e^{\tilde{a}\Phi}F^{M_{1}...M_{(D-n)}}] = 0 \nonumber
\end{eqnarray}
for the magnetic ans\H{a}tz.
Now the rest of the procedure to find electric/magnetic $p$-brane solutions consists
simply of writing and solving the field equations given above in terms of the
$SO(D-d)$-symmetric ans\H{a}tz of the fields given. And central to this procedure is the
computation of Ricci tensor components $R_{MN}$ which, as stated earlier, can be done most
easily in the context of Riemann-Cartan formulation in which the line element is given
in non-coordinate orthonormal basis as in eq.(4). Namely, one first obtains the spin
connection 1-form $\omega^{I}_{J}=\omega^{I}_{MJ}dX^{M}$ via the Cartan's 1st structure
equation
\begin{eqnarray}
de^{I} + \omega^{I}_{J}\wedge e^{J} = 0
\end{eqnarray}
and from them, one next calculates the Riemann curvature 2-form 
$R^{IJ}=(1/2)R^{IJ}_{MN}dX^{M}\wedge dX^{N}$ via the Cartan's 2nd structure equation
\begin{eqnarray}
R^{IJ} = d\omega^{IJ} + \omega^{IK}\wedge \omega_{K}^{J}
\end{eqnarray}
to get, upon projecting back in the coordinate basis,
\begin{eqnarray}
R_{\mu\nu} &=& \hat{R}_{\mu\nu} - \gamma_{\mu\nu}(x)e^{2(A-B)}\left[A'' + {(\tilde{d} +1)
\over r}A' + d(A')^2 + \tilde{d}A'B'\right], \nonumber \\
R_{mn} &=& -\delta_{mn}\left[B'' + d A'B' + \tilde{d}(B')^2 + {(2\tilde{d}+1)\over r}B'
+ {d\over r}A'\right] \\
&& - {y^{m}y^{n}\over r^2}\left[\tilde{d}B'' + d A'' - 2d A'B' + d(A')^2 - \tilde{d}(B')^2
- {\tilde{d}\over r}B' - {d\over r}A'\right], \nonumber \\
R_{\mu m} &=& 0 \nonumber
\end{eqnarray}
where prime denotes the derivative with respect to $r=(y^{m}y^{m})^{1/2}$ and in the first line,
$\hat{R}_{\mu\nu}$ denotes the Ricci tensor associated with the non-trivial metric on the brane 
$\gamma_{\mu\nu}(x)$.
Finally, the field equations written in terms of $SO(D-d)$-symmetric ans\H{a}tz of the fields
take the forms
\begin{eqnarray}
&&\hat{R}_{\mu\nu} - \gamma_{\mu\nu}(x)e^{2(A-B)}\left[A'' + {(\tilde{d} +1)\over r}A' + d(A')^2 
+ \tilde{d}A'B'\right] = -\gamma_{\mu\nu}(x)e^{2(A-B)}{\tilde{d}\over 2(D-2)}S^2, \nonumber \\
&&B'' + d A'B' + \tilde{d}(B')^2 + {(2\tilde{d}+1)\over r}B' + {d\over r}A' = -{d\over 2(D-2)}S^2,
\\
&&\tilde{d}B'' + d A'' - 2d A'B' + d(A')^2 - \tilde{d}(B')^2 - {\tilde{d}\over r}B' - {d\over r}A'
+ {1\over 2}(\Phi')^2 = {1\over 2}S^2 \nonumber
\end{eqnarray}
for the metric field equations and
\begin{eqnarray}
\Phi'' + (dA'+\tilde{d}B')\Phi' + {(\tilde{d}+1)\over r}\Phi' = {1\over 2}\zeta \tilde{a}S^2
\end{eqnarray}
for the dilaton field equation. And here we denoted by $S$ the quantity
\begin{eqnarray}
S &\equiv& \left[e^{(\tilde{a}\Phi/2-dA+C)}C'\right] ~~~({\rm electric ~case} : d=(n-1), ~\zeta =-1), \\
S &\equiv& \left[\lambda e^{(\tilde{a}\Phi/2-\tilde{d}B)}{1\over r^{(\tilde{d}+1)}}\right]
~~~({\rm magnetic ~case} : \tilde{d}=(D-n-1), ~\zeta =1). \nonumber
\end{eqnarray}
Lastly, we note that the remaining antisymmetric R-R tensor gauge field equation
(i) for the electric case reads
\begin{eqnarray}
&&\left[C'' + {(\tilde{d}+1)\over r}C'\right] + C'(C' + \tilde{a}\Phi' - dA' + \tilde{d}B') = 0, \\
&&e^{(\tilde{a}\Phi-dA+\tilde{d}B)}(\partial_{m}e^{C})\epsilon_{\nu \nu_{1}...\nu_{(n-2)}}\partial_{\mu}
[(det ~\gamma(x))\gamma^{\mu\nu}\gamma^{\mu_{1}\nu_{1}}...\gamma^{\mu_{(n-2)}\nu_{(n-2)}}] = 0
\nonumber
\end{eqnarray}
whereas (ii) for the magnetic case, the R-R tensor gauge field equation always holds regardless 
of the specific solution forms for the metric and the dilaton fields. \\
We now attempt to solve these coupled field equations. The field equations at hand, however,
look quite formidable yet. To proceed any further, therefore, we need some hint from the
requirements for supersymmetry preservation. Namely, we shall now refine the solution ans\H{a}tz
above by imposing the ``linearity'' condition 
\begin{eqnarray}
dA' + \tilde{d}B' = 0.
\end{eqnarray}
This condition, as one can readily see in the field equations above, amounts to eliminating 
$B(r)$ in favor of $A(r)$ in the metric and dilaton field equations where it is interesting,
even at this early stage, to note that these field equations themselves require the
``Ricci-flatness'' of the $p$-brane worldvolume metric $\gamma_{\mu\nu}(x)$. That is, due
to the imposed condition $dA' + \tilde{d}B' = 0$, in order for the first and second equations
of the metric field equation in (13) to be consistent, it is required that $\hat{R}_{\mu\nu}=0$.
Next, further refining our solution ans\H{a}tz by setting 
\begin{eqnarray}
\Phi' = {\zeta \tilde{a} (D-2)\over \tilde{d}}A'
\end{eqnarray}
the remaining set of three coupled equations for $A(r)$ and $\Phi(r)$, two from eq.(13) and one in
eq.(14), can be readily integrated
to yield
\begin{eqnarray}
e^{2A} &=& \exp{\left[{2\tilde{d}\over \zeta \tilde{a}(D-2)}\Phi\right]} = H^{-4\tilde{d}/\Delta(D-2)}(r), 
\nonumber \\
e^{2B} &=& \exp{\left[-{2d\over \zeta \tilde{a}(D-2)}\Phi\right]} = H^{4d/\Delta(D-2)}(r), ~~~{\rm with}
\nonumber \\
&&H(r) = 1 + {k\over r^{\tilde{d}}}
\end{eqnarray}
being the harmonic function as a solution to the Laplace equation in $(D-d)=(\tilde{d}+2)$-spacelike
transverse dimensions. The integration constant $k$ sets the mass scale of the solution and has been
taken to be positive to ensure the absence of naked singularities at finite $r$.
And here as for the other integration constants, $\Phi(\infty)$ has been set to zero for
simplicity while we have chosen $A(\infty)=0=B(\infty)$ so that the solution tends to flat space
at transverse infinity $r\rightarrow \infty$. Also note that we defined a parameter $\Delta$ such that
\begin{eqnarray}
\tilde{a}^2 = \Delta - {2d\tilde{d}\over (D-2)}.
\end{eqnarray}
Now, what remains is to check if these metric and dilaton solutions obtained are really consistent
with the R-R tensor gauge field equations. As we already noted earlier, (ii) for the magnetic case,
the tensor gauge field equations always hold regardless of the specific solution forms for the
metric and dilaton fields. And this implies that the supergravity field equations generally admit
``Ricci-flat'' {\it magnetic} $p$-brane solutions. Next, (i) for the electric case,
the first of the set of two R-R tensor gauge field equations given in eq.(16) is actually implied
by the metric and dilaton field equations and hence can be readily integrated to yield 
\begin{eqnarray}
e^{C} = {2\over \sqrt{\Delta}}H^{-1}(r).
\end{eqnarray}
And it is straightforward to see that the second of the two R-R tensor gauge field equations
trivially holds. Thus to conclude, both the electric and magnetic Ricci-flat $p$-branes turn
out to be legitimate supergravity solutions. For $p\leq 2$, however, there are no non-trivial
Ricci-flat $p$-brane solutions since a Ricci-flat manifold with spacetime dimensions less than 
or equal to 3 is necessarily flat.\footnote{We would like to thank D. Brecher (private communication)
for pointing out an incorrect statement regarding the electric Ricci-flat $p$-brane solutions
in the earlier version of the present work.}  Finally, we summarize the
dilatonic {\it Ricci-flat} $p$-brane solutions of general supergravity theories as
\begin{eqnarray}
ds^2 &=& H^{-4\tilde{d}/\Delta(D-2)}(r)[\gamma_{\mu\nu}(x)dx^{\mu}dx^{\nu}] + 
 H^{4d/\Delta(D-2)}(r)[dr^2 + r^2d\Omega^2_{(\tilde{d}+1)}], \nonumber \\
e^{\Phi(r)} &=& H^{-2\tilde{a}/\zeta \Delta}(r) ~~~{\rm with} \\
&&\hat{R}_{\mu\nu}(\gamma) = 0, ~~~H(r)=1 + {k\over r^{\tilde{d}}} \nonumber
\end{eqnarray}
where we introduced the spherical-polar coordinates for the transverse dimensions
$dy^{m}dy^{m} = dr^2 + r^2d\Omega^2_{(\tilde{d}+1)}$ with $d\Omega^2_{(\tilde{d}+1)}$
being the metric on unit $(\tilde{d}+1)$-sphere. 
And (i) for the electric case ($\zeta = -1$), the R-R tensor gauge field strength is
given by
\begin{eqnarray}
F_{m\mu_{1}...\mu_{n-1}} = \epsilon_{\mu_{1}...\mu_{n-1}}\partial_{m}
\left[{2\over \sqrt{\Delta}}H^{-1}(r)\right]
\end{eqnarray}
whereas (ii) for the magnetic case ($\zeta = 1$), the R-R tensor gauge field strength is
\begin{eqnarray}
F_{m_{1}m_{2}...m_{(D-n)}} = \lambda \epsilon_{m_{1}m_{2}...m_{(D-n)}l}
{y^{l}\over r^{(D-n+1)}} ~~~{\rm with} ~~~k={\sqrt{\Delta}\over 2\tilde{d}}\lambda.
\end{eqnarray}

{\rm \bf 2. Applications}

Consider now, as a particular example, the bosonic sector of $D=11$ SUGRA represented by
the action
\begin{eqnarray}
S_{11} &=& \int d^{11}X \sqrt{G}[R - {1\over 2\cdot 4!}F^2_{[4]}] - {1\over 6}\int
A_{[3]}\wedge F_{[4]}\wedge F_{[4]} ~~~{\rm where} \nonumber \\
A_{[3]} &=& {1\over 3!}A_{[MNP]}dX^{M}\wedge dX^{N}\wedge dX^{P}, ~~~F_{[4]}=dA_{[3]}.
\end{eqnarray}
When we attempt to apply the results associated with the Ricci-flat dilatonic $p$-brane
solutions obtained above to this $D=11$ SUGRA system, we first need to note two
particular points. The first is that, here, no scalar field is present. This follows
from the supermultiplet structure of $D=11$ SUGRA theory in which all fields are gauge
fields. In lower dimensions, of course, scalars do emerge ; namely the dilaton field
in $D=10$ type IIA SUGRA appears upon dimensional reduction from $D=11$ SUGRA to
$D=10$. The absence of the scalar in $D=11$ SUGRA can then be handled, in the context of
our general discussion provided above, by simply identifying the dilaton coupling parameter
``$\tilde{a}$'' with zero so that the scalar may be consistently truncated.
Namely, from $\tilde{a}^2 = \Delta - 2d\tilde{d}/(D-2)$, we identify $\Delta=2\cdot 3\cdot 6/9
=4$ for the $D=11$ SUGRA case. The second point to note is the presence of $AFF$ Chern-Simons
term in the action. This term is required by $D=11$ local supersymmetry with the coefficient
precisely as given. Although we did not take into account the effects of this $AFF$ term in 
our general discussion, this omission is not essential to the basic class of $p$-brane
solutions we are considering. \\
Now, as an obvious example of Ricci-flat, magnetic $M$-brane solution to $D=11$ SUGRA,
we consider the $M5$-brane solution. Plugging $D=11$, $\tilde{a}=0$, $d=(p+1)=6$, 
$\tilde{d}=3$, and $\Delta=4$ in the general solution given in eqs.(22) and (24), it is given by
(also see [11])
\begin{eqnarray}
ds^2 &=& H^{-1/3}(r)[\gamma_{\mu\nu}(x)dx^{\mu}dx^{\nu}] +
 H^{2/3}(r)[dr^2 + r^2d\Omega^2_{4}], \nonumber \\
H(r) &=& 1 + {k\over r^3},  ~~~{\rm and}  \\
F_{m_{1}m_{2}m_{3}m_{4}} &=& \epsilon_{m_{1}m_{2}m_{3}m_{4}n}{3ky^{n}\over r^5} \nonumber
\end{eqnarray}
where we used $\lambda = 2\tilde{d}k/\sqrt{\Delta} = 3k$ and $\mu,~\nu = 0,...5$ and 
$m,~n = 6,...10$. At first glance, it may look like that the metric/gauge solutions are
singular at $r=0$ just as we encountered for the Poincare-invariant $M5$-brane case.
However, if one evaluates invariant components of the curvature tensor and the R-R gauge
field strength, one readily finds that these invariants are non-singular at $r=0$. 
Moreover, although the proper distance to the surface $r=0$ along a $t=const.$ geodesic
diverges, the surface $r=0$ can be reached along null geodesics in finite affine parameter.
This implies that the timelike hypersurface $r=0$ is a horizon, i.e., a coordinate 
singularity. Besides, close inspection reveals that actually $r=0$ is a ``degenerate''
horizon which is known to occur typically for extremal black holes when the inner and
outer horizons coalesce. The near-horizon geometry of this $M5$-brane metric solution, 
however, fails to be that of $AdS_7 \times S_4$ due to the Ricci-flat nature of the
worldvolume geometry. And the integration constant $k$ appearing in the Harmonic function
$H(r)$ representing the mass scale of the solution can be explicitly determined for
Poincare-invariant worldvolume metric solution $\gamma_{\mu\nu}(x)=\eta_{\mu\nu}$ in which
case, $k=\pi N l^3_{11}$ with $l_{11}$ denoting the 11-dimensional Planck length and
$N$ the number of branes generally for stack of coincident branes. \\ 
We now turn to the bosonic sector of $D=10$ type IIA/IIB SUGRA theories. It is 
well-recognized that $D=10$ type IIA/IIB SUGRA theories can be thought of as the low
energy effective theories of type IIA/IIB superstring theories. And in the application
of our general $p$-brane solutions given earlier to this $D=10$ type IIA/IIB SUGRA theories,
solutions of particular interest are the ``$\tilde{a}=0$ subsets'' for which the dilaton is set
to zero or just a constant. Thus here, we just consider these ``non-dilatonic'' $p$-brane
solutions. \\
Our first example is the Ricci-flat, magnetic $D3$-brane solution of type IIB theory. 
Plugging $D=10$, $\tilde{a}=0$, $d=(p+1)=4$, $\tilde{d}=4$, and $\Delta=4$ in eqs.(22) and (24), 
it is given by
\begin{eqnarray}
ds^2 &=& H^{-1/2}(r)[\gamma_{\mu\nu}(x)dx^{\mu}dx^{\nu}] +
 H^{1/2}(r)[dr^2 + r^2d\Omega^2_{5}], \nonumber \\
H(r) &=& 1 + {k\over r^4},  ~~~{\rm and}  \\
F_{m_{1}...m_{5}} &=& \epsilon_{m_{1}...m_{5}n}{4ky^{n}\over r^6} \nonumber
\end{eqnarray}
where we used $\lambda = 2\tilde{d}k/\sqrt{\Delta} = 4k$ and $\mu,~\nu = 0,...3$ and
$m,~n = 4,...9$. The 5-form R-R tensor field strength here is self-dual.
Again, the $r=0$ timelike hypersurface is a harmless degenerate horizon 
and the integration constant $k$ appearing in the Harmonic function $H(r)$ can be explicitly
determined for Poincare-invariant worldvolume metric solution $\gamma_{\mu\nu}(x)=\eta_{\mu\nu}$
in which case, $k=4\pi g_{s}N l^2_{s}$ with $g_{s}$ being the string coupling constant and
$l_{s}=\sqrt{\alpha'}$ the string length. And due to the generally Ricci-flat nature of the
worldvolume metric, the near-horizon geometry of this $D3$-brane fails to be that of 
$AdS_{5}\times S^{5}$. \\
Lastly, we consider the Ricci-flat, magnetic $D4$-brane solution of type IIA theory which is 
of our central concern in this work. Plugging $D=10$, $\tilde{a}=0$, $d=(p+1)=5$, $\tilde{d}=3$, 
and $\Delta=15/4$ in eqs.(22) and (24), it is given by
\begin{eqnarray}
ds^2 &=& H^{-2/5}(r)[\gamma_{\mu\nu}(x)dx^{\mu}dx^{\nu}] +
 H^{2/3}(r)[dr^2 + r^2d\Omega^2_{4}], \nonumber \\
H(r) &=& 1 + {k\over r^3},  ~~~{\rm and}  \\
F_{m_{1}...m_{4}} &=& \epsilon_{m_{1}...m_{4}n}{12\over \sqrt{15}}{ky^{n}\over r^5} \nonumber
\end{eqnarray}
where we used $\lambda = 2\tilde{d}k/\sqrt{\Delta} = 12k/\sqrt{15}$ and $\mu,~\nu = 0,...4$ and
$m,~n = 5,...9$. $r=0$ is again a degenerate horizon and for  Poincare-invariant worldvolume metric 
solution, the integration constant $k$ can be explicitly determined to be 
$k \sim g_{s}N l^3_{s}$.  Here, also the generally Ricci-flat nature of the worldvolume keeps the
near-horizon geometry of this $D4$-brane from being that of $AdS_{6}\times S^{4}$. \\
We now end with the remark on the asymptotic geometry of this $D4$-brane metric solution as
$r\rightarrow \infty$ [11]. Far from the source $4$-brane, or more precisely, at infinities in
null directions, $H(r)\rightarrow 1$ and hence
\begin{eqnarray}
ds^2 \simeq \gamma_{\mu\nu}(x)dx^{\mu}dx^{\nu} + dy^{m}dy^{m}
\end{eqnarray}
which is Ricci-flat as a whole. For instance, therefore, we may take the Ricci-flat
worldvolume metric as that of (see also [11])
\begin{eqnarray}
&&R \times ({\rm Taub-NUT}) : ~~~ds^2 = [-dt^2 + d\hat{s}^2_{TN}] + dy^{m}dy^{m}, \\
&&R \times ({\rm Eguchi-Hanson}) : ~~~ds^2 = [-dt^2 + d\hat{s}^2_{EH}] + dy^{m}dy^{m} \nonumber
\end{eqnarray}
where $d\hat{s}^2_{TN}$ and $d\hat{s}^2_{EH}$ denote the  Taub-NUT [18] and Eguchi-Hanson [19]
``gravitational instantons'' satisfying the Ricci-flat 4-dimensional worldspace metric
and obeying (anti) self-dual worldspace Riemann curvature tensor. \\
Now, this completes the construction of Ricci-flat, extremal $p$-brane solutions in 
supergravity theories.
Eventually, we shall be interested in the explicit construction of $SU(2)$ Yang-Mills instanton
solutions in the background geometry of a stack of two coincident probe $D4$-brane worldspaces
particularly when
the metric of target spacetime in which the probe branes are embedded is given by the
Ricci-flat, magnetic extremal 4-brane solution in type IIA theory with its worldspace metric
being given by that of Taub-NUT or Eguchi-Hanson gravitational instanton. And then we shall
demonstrate that with this YM instanton-gravitational instanton configuration on the probe
$D4$-branes' worldvolume, the energy of the probe branes attains its minimum value and hence
enjoys stable state provided one employs the Tseytlin's non-abelian DBI
action [4] for the description of multiple probe $D$-branes.

\begin{center}
{\rm \bf III. Non-abelian DBI action which ``knows'' energy-minimizing \\
worldvolume solitons}
\end{center}

In general, the abelian Dirac-Born-Infeld (DBI) action describing the low energy 
dynamics of a single probe $Dp$-brane is given by [1]
\begin{eqnarray}
S^{A}_{DBI} = {1\over (2\pi)^{p}(\alpha')^{p+1\over 2}}\int d^{p+1}x
e^{-\Phi } \sqrt{det ~[g_{\mu\nu}+2\pi \alpha' (B+F)_{\mu\nu}]} 
\end{eqnarray}
where 
$T_{p}\equiv (1/ (2\pi)^{p}(\alpha')^{(p+1)/2})$ is the brane tension and inside the
square root, $|det ~[g_{\mu\nu}+2\pi \alpha' (B+F)_{\mu\nu}]|$, i.e., its absolute value is understood. 
$\Phi$ is the dilaton field and 
$B_{\mu\nu}$ and $g_{\mu\nu}$ are the ``pullbacks'' of bulk NS-NS antisymmetric tensor gauge field and 
the target spacetime metric to the probe brane respectively in the sense, say, that
\begin{eqnarray}
g_{\mu\nu} = G_{MN}{\partial X^{M}\over \partial x^{\mu}}
{\partial X^{N}\over \partial x^{\nu}}|_{brane} 
= G_{\mu\nu} + \partial_{\mu}\phi^{m}\partial_{\nu}\phi^{m} 
\end{eqnarray}
where again $\mu,~\nu = 0,1,...p$ ~and $m = (p+1),...(D-1)$ ~denote the indices for longitudinal and
transverse coordinates to the brane as before and we identified the ``worldvolume scalars'' 
as $\phi^{m}\equiv y^{m}$ 
representing the excitations of the brane along directions transverse to the brane worldvolume. 
Hereafter, however, we shall only consider the case when no worldvolume scalars are excited, i.e.,
$\phi^{m}=0$ for all $m=(p+1),...(D-1)$. Thus one may assume $g_{\mu\nu} = G_{\mu\nu}$ throughout. 
$F_{\mu\nu}$ is the $U(1)$ gauge field strength living on the brane that represents the 
fluctuations of the brane along longitudinal directions. Thus it should not be confused with
the R-R tensor gauge field strength in the previous section. Note also 
that in this expression for the DBI action above, we fixed the worldvolume reparametrization
invariance by taking the physical ``static'' gauge in which the worldvolume parameters
$\{\sigma^{\mu}\}$ are identified with the first $(p+1)$ target spacetime 
coordinates
\begin{eqnarray}
\sigma^{\mu} = x^{\mu}, ~~~~(\mu=0,...p).
\end{eqnarray}
We now consider the target spacetime with geometry given by that of the $p$-brane solution 
arising in various supergravity theories such as the $D=11$ SUGRA or $D=10$ type IIA/IIB 
(non-dilatonic) SUGRA. And particularly, consider the source $p$-brane solutions as the
``Ricci-flat'' solutions we discussed in the previous section given by
\begin{eqnarray}
ds^2 &=& G_{MN}dX^{M}dX^{N} \\
&=& H^{-{4\tilde{d}/\Delta (D-2)}}(r)[\gamma_{\mu\nu}(x)dx^{\mu}dx^{\nu}] +
 H^{{4d/\Delta (D-2)}}(r)dy^{m}dy^{m} \nonumber \\
{\rm with} ~~~&&H(r) = 1 + {k\over r^{\tilde{d}}}.
\end{eqnarray}
Namely, we consider a test probe $Dp$-brane in the background of Ricci-flat source $p$-brane
geometry.
Then the pullback of this target spacetime metric $G_{MN}$ to the probe $Dp$-brane worldvolume
is 
\begin{eqnarray}
g_{\mu\nu} = G_{MN}{\partial X^{M}\over \partial x^{\mu}}{\partial X^{N}\over \partial x^{\nu}}
=  H^{-{4\tilde{d}/\Delta (D-2)}}(r_{0})\gamma_{\mu\nu}(x).
\end{eqnarray}
Further, we may take a (non-coordinte) orthonormal basis for this $(p+1)$-dimensional $Dp$-brane
worldvolume, i.e., $e^{a}=e^{a}_{\mu}dx^{\mu}$ such that $g_{\mu\nu}=\eta_{ab}e^{a}_{\mu}e^{b}_{\nu}$. 
Then the tetrad components of the worldvolume gauge field is
\begin{eqnarray}
F_{ab} = F_{\mu\nu}e_{a}^{\mu}e_{b}^{\nu},  ~~~{\rm or ~~inversly}
~~~F_{\mu\nu} = F_{ab}e^{a}_{\mu}e^{b}_{\nu}
\end{eqnarray}
then obviously
\begin{eqnarray}
F^2 = F_{\mu\nu}F^{\mu\nu} &=&  F_{ab}F^{ab}, ~~~F\cdot \tilde{F} = F_{\mu\nu}\tilde{F}^{\mu\nu} 
= F_{ab}\tilde{F}^{ab}, \nonumber \\
\tilde{F}^2 &=& \tilde{F}_{\mu\nu}\tilde{F}^{\mu\nu} =  \tilde{F}_{ab}\tilde{F}^{ab},
~~~{\rm etc.}
\end{eqnarray}
Particularly, therefore, if we turn off the dilaton $\Phi (x)$ and the NS-NS gauge field
$B_{\mu\nu}(x)$, the DBI action, in terms of the tetrad components of the tensor fields involved,
takes the form
\begin{eqnarray}
S^{A}_{DBI} &=& T_{p}\int d^{p+1}x [\sqrt{g} - \sqrt{det ~(g_{\mu\nu}+F_{\mu\nu})}] \\
&=& T_{p}\int d^{p+1}x \sqrt{g}[1 - \sqrt{det ~(\eta_{ab}+F_{ab})}] \nonumber
\end{eqnarray}
for a {\it single} probe $Dp$-brane with abelian worldvolume gauge field. Note that here we 
absorbed the factor $2\pi \alpha'$ into the redefinition of $F_{\mu\nu}$ and
added a constant term corresponding to the Lagrangian density $\sim T_{p}\sqrt{g}$ to the
DBI action and used $det ~(g_{\mu\nu}+F_{\mu\nu}) = det ~(e^{a}_{\mu}e^{b}_{\nu})det ~(\eta_{ab}+F_{ab})
=g ~det ~(\eta_{ab}+F_{ab})$. \\
We now generalize our discussion to the case of a stack of $N$-coincident probe $Dp$-branes.
Then following the argument of Witten [3], non-abelian $SU(N)$ gauge theory should provide a
good description of the relevant low energy dynamics of $N$-coincident $Dp$-branes and hence
the non-abelian generalization of a multiple $Dp$-brane DBI action with $SU(N)$ worldvolume
gauge field may be taken as
\begin{eqnarray}
S^{A}_{DBI} = T_{p}\int d^{p+1}x \sqrt{g}
\{Tr, ~~Str, ~~(Str + i~Atr)\}[I - \sqrt{det ~(\eta_{ab}+F_{ab})}]
\end{eqnarray}
where $I$ is the unit $SU(N)$ matrix and now $F_{ab}=F^{A}_{ab}T^{A}$ with $T^{A}$
$(A=1,2,...(N^2-1))$ being the $SU(N)$ generators. Of course, here, the ``traces'' are
over the group indices to make the DBI action a group scalar and we defined
\begin{eqnarray}
Str(M_{1}...M_{n}) &\equiv& {1\over n!}\Sigma_{\pi}Tr(M_{\pi(1)}...M_{\pi(n)}), \\
Atr(M_{1}...M_{n}) &\equiv& {1\over n!}\Sigma_{\pi}(-1)^{\pi}Tr(M_{\pi(1)}...M_{\pi(n)})
\nonumber
\end{eqnarray}
and the factor ``$i$'' in $(Str + i~Atr)$ is introduced since the basis of the group
algebra is Hermitian. And among the different choices for the trace operations, 
$Tr[...]$ can be found in Polchinski's review article on $D$-branes [1], $Str[...]$ was
proposed by Tseytlin [4], and lastly $(Str + i~Atr)[...]$ has been suggested by
Argyres and Nappi [5]. And in the most general sense, concerning the question as to what
the correct generalization of the DBI action to the non-abelian case is (i.e., which
specific trace operation one should take), the issue seems to be somewhat ambiguous. 
Nevertheless, here it will be argued that the $Str[...]$ proposal put forward
by Tseytlin is indeed singled out by noting that with this choice, the non-abelian DBI
action ``knows'' about {\it energy-minimizing BPS states}, or worldvolume solitons.
The following argument is taken directly from [10,11] and as was pointed out there,
it can be made explicit by taking the $D4$-brane for example.
Note that for $D4$-brane case, the worldvolume is $(p+1)=5$-dimensional and hence the
spacetime determinant of the DBI action can be computed as [10]
\begin{eqnarray}
|det ~(\eta_{ab}+F_{ab})| &=& - det ~(\eta_{ab}I+F_{ab}) \\
&=& I + {1\over 2}F^2 + {1\over 3}F^3 - {1\over 4}[F^4-{1\over 2}(F^2)^2] + {1\over 5}
F^5 + {1\over 12}(F^2F^3 + F^3F^2) \nonumber
\end{eqnarray}
where $F^2 = F_{ab}F^{ab}$, $F^3 = F_{ab}F^{bc}F_{c}^{a}$, $F^4 = F_{ab}F^{bc}F_{cd}F^{da}$,
and $F^5 =  F_{ab}F^{bc}F_{cd}F^{de}F_{e}^{a}$. In the abelian case, all the odd powers of
$F_{ab}$ vanish but this is not true for the case at hand. Besides, since the complete DBI
action involves $[- det ~(\eta_{ab}I+F_{ab})]^{1/2}$, we need to evaluate the binomial 
expansion of above expression which would obviously results in an infinite series containing
terms of both even and odd powers of $F_{ab}$. And the trace operation should be taken ``after''
this binomial series expansion. Then the important properties of the $Str$ and $Atr$ operations
is that they pick out just the even and odd powers of $F_{ab}$ respectively. This point is
clear if one takes the $SU(2)$ case for example and moreover, at least to the first few
orders, the same will apply to the cross terms generated by the binomial expansion ;
that is, e.g., $Str(F^2F^3) = Str(F^2F^5) = 0$. Generally speaking, therefore, the non-abelian
version of DBI action with ``$Str$'' can be written as a sum of even powers of $F_{ab}$ alone,
whereas that with ``$Tr$'' or ``$Atr$'' would involve odd powers of $F_{ab}$ as well.
Indeed this point was the motivation behind Tseytlin's proposal. Since odd powers of $F_{ab}$
can be written in terms of derivatives of $F_{ab}$, i.e.,
$F^3 \sim [F,F]F \sim [D,D]F$, they, Tseytlin thought, should not appear in non-abelian DBI
theory action just as the abelian DBI action does not involve derivatives of the field strength.
Thus Tseytlin was led to define the non-abelian DBI action in terms of ``$Str$'' operation alone
so that it depends only on even powers of $F_{ab}$. Having reviewed the nature of Tseytlin's
proposal for the non-abelian DBI action, we now provide an evidence that strongly supports
Tseytlin's action by invoking the argument concerning the energy-minimizing BPS states or
worldvolume solitons. Here, in this work, we are particularly interested in the ``instantons
in $D4$-branes'' and thus we consider static configurations of $D4$-branes with no worldvolume
scalars being excited. Then $F_{\alpha 0} = E_{\alpha} = 0$ (where $a,b=0,1,...p=4$ are 
worldvolume and $\alpha, \beta = 1,...4$ are worldspace indices respectively) and hence
\begin{eqnarray}
- det ~&&(\eta_{ab}I+F_{ab}) = - (\eta_{00})det ~(\eta_{\alpha\beta}I+F_{\alpha\beta}) 
\nonumber \\
&&= 1 + {1\over 4}F_{\alpha\beta}F^{\alpha\beta} + {1\over 4}\tilde{F}_{\alpha\beta}
\tilde{F}^{\alpha\beta} + {1\over 16}(F_{\alpha\beta}\tilde{F}^{\alpha\beta})^2 \\
&&= \left(1 \pm {1\over 4}F_{\alpha\beta}\tilde{F}^{\alpha\beta}\right)^2 +
{1\over 4}\left(F_{\alpha\beta} \mp \tilde{F}_{\alpha\beta}\right)^2. \nonumber
\end{eqnarray}
Note, here, that $\tilde{F}_{\alpha\beta}=\epsilon_{\alpha\beta\lambda\sigma}F^{\lambda
\sigma}/2$ is the Hodge dual of $F_{\alpha\beta}$ with respect to the worldspace
indices only. Therefore, Tseytlin's non-abelian DBI action for the static configurations 
of a $D4$-brane becomes
\begin{eqnarray}
L_{DBI} &=& T_{4}\sqrt{g} Str[I - \sqrt{det ~(\eta_{ab}+F_{ab})}] \\
&=& T_{4}\sqrt{g} Str\left[I - \sqrt{\left(I \pm {1\over 4}F\cdot \tilde{F}\right)^2 +
{1\over 4}|F \mp \tilde{F}|^2}\right] \nonumber 
\end{eqnarray}
where to get the expression in the second line, use has been made of the symmetry properties
of the ``$Str$'' operation. That is, since the symmetric trace will be taken only at the
end, during the course of calculation, the matrices $F_{\alpha\beta}=(F^{A}_{\alpha\beta}T^{A})$
can be treated as if they were ``abelian'' until the last at which point the non-commuting
group generators can be re-inserted. And of course, this scheme can be justified only if we
employ Tseytlin's ``$Str$'' proposal in which one can effectively assume $AB=BA$. Obviously,
therefore, one would never get the expression in the second line above for the non-abelian
DBI action if he or she employed other trace operations such as ``$Tr$'' or ``$Atr$''.
Then what is so special about this last expression for the non-abelian DBI action ?
Notice that for the (anti) self-dual non-abelian gauge field configuration having
$F_{\alpha\beta} = \pm \tilde{F}_{\alpha\beta}$, the determinant can be written as a
``complete square'' {\it linearizing} the DBI action and then turning it into that of usual
Yang-Mills theory. What is more, since we are dealing with static configurations, the
energy density of the $D4$-brane is just $H_{DBI} = - L_{DBI}$, which gets {\it minimized}
if and only if $F_{\alpha\beta} = \pm \tilde{F}_{\alpha\beta}$. Thus the BPS condition,
or the (anti) self-duality of non-abelian gauge theory instanton solution at once linearizes
the otherwise highly non-linear DBI action and minimizes the energy of the $D4$-brane provided
Tseytlin's non-abelian DBI action is employed [10,11]. \\
Before closing, a cautious comment might be relevant to eliminate a possible confusion.
It might seem as if we simply put the non-abelian gauge field configuration having
$F_{\alpha\beta} = \pm \tilde{F}_{\alpha\beta}$ {\it by hand} into  $L_{DBI} = - H_{DBI}$ 
and then announce naively that, if one does so, the DBI action linearizes and the Hamiltonian
gets minimized. What happens, however, is that the YM gauge field having 
$F_{\alpha\beta} = \pm \tilde{F}_{\alpha\beta}$, or the BPS condition, is indeed a solution
to the Euler-Lagrange's equation of motion that results from the DBI action provided one
employs the Tseytlin's action much as the same BPS condition automatically implies the
linear gauge field equation in ordinary YM gauge theory. To see this, consider that generally
the (non-abelian) gauge field $A_{\mu}$ ($\mu = 0,...4$) living on the worldvolume and
the worldvolume scalar $\phi^{m}=y^{m}$ ($m=5,...9$) represent the excitations of the probe
$D4$-brane along directions longitudinal and transverse to the worldvolume. Thus in principle,
both $A_{\mu}$ and $\phi^{m}$ are to be dynamically determined by solving the Euler-Lagrange's
equation of motion that can be obtained by extremizing the DBI action. Now the expression
for the DBI action for a probe $D4$-brane given in eq.(44) holds for the case with no
worldvolume scalars and static configurations (provided the Tseytlin's action is adopted)
and manifestly it gets {\it extremized} for $F_{\alpha\beta} = \pm \tilde{F}_{\alpha\beta}$.
Namely, the non-abelian gauge field configuration having the (anti) self-dual field
strength is itself the saddle point of the DBI action satisfying the associated 
Euler-Lagrange's equation of motion. Thus in order to demonstrate that this procedure
can actually work, the actual question would be whether or not one can construct, say,
$SU(2)$ Yang-Mills instanton solution having (anti) self-dual field strength in the
background of {\it Ricci-flat} $D4$-brane worldspace with metric given, say, by 4-dimensional
gravitational instantons. The answer is, as we shall see in the next section, ``yes''. 

\begin{center}
{\rm \bf IV. YM instantons in gravitational instanton backgrounds}
\end{center}
 
 Here in this section, eventually in order to construct the SU(2) YM instanton solutions
 particularly in the background of TN and EH gravitational instantons, 
 we begin by presenting a ``simply physical'' and hence perhaps the most direct
 algorithm for generating the YM instanton solutions practically in all species
 of known GI [16]. As we shall see in a moment, the essence of this method lies 
 in writing the (anti) self-dual YM equation by employing truly relevant ans\H{a}tz for the
 YM gauge connection and then directly solving it. After presenting the general formulation
 describing the algorithm, we shall apply the algorithm to the case of the TN 
 and the EH metrics, the two best-known GI.  As we shall discuss later on in the
 Appendix, 
 interestingly the solutions to (anti) self-dual YM equation turn out to be the rather 
 exotic type of instanton configurations which are everywhere non-singular having {\it finite}
 YM action but sharing some features with meron solutions [20] such as 
 their typical structure and generally {\it fractional} topological  charge values
 carried by them. Namely, the YM instanton solution that we shall discuss in
 the background of GI in this work exhibit characteristics which are mixture
 of those of typical instanton and typical meron. This seems remarkable 
 since it is well-known that in flat spacetime, meron does not solve the
 1st order (anti) self-dual equation although it does the 2nd order YM
 field equation and is singular at its center and has divergent action. \\  
 In the loose sense, GI may be
 defined as a positive-definite metrics $g_{\mu\nu}$ on a complete
 and non-singular manifold satisfying the Euclidean Einstein
 equations and hence constituting the stationary points of the
 gravity action in Euclidean path integral for quantum gravity.
 There are several solutions to Euclidean Einstein equations that can fall 
 into the category of the GI of this sort.
 But in the stricter sense [21], they are the metric solutions to the
 Euclidean Einstein equations having (anti) self-dual Riemann
 tensor
\begin{eqnarray}
\tilde{R}_{abcd} = {1\over 2}\epsilon_{ab}^{~~ef} R_{efcd} = \pm
R_{abcd}
\end{eqnarray}
(say, with indices written in non-coordinate orthonormal basis)
and include only two families of solutions in a rigorous sense ;
the TN metric [18] and the EH instanton [19]. 
Recall that we are mainly interested in the explicit construction of SU(2) YM instanton
solutions in the background of a probe $D4$-brane worldspace geometry particularly when
the metric of target spacetime in which the probe brane is embedded is given by the
Ricci-flat, magnetic extremal 4-brane solution in type IIA theory with its worldspace metric
being given by that of TN or EH GI. Thus in this section, we shall be
interested exclusively in the construction of YM instantons in the background of these 
two GI satisfying the rigorous definition. \\
We now begin with the action governing our system, i.e., the Einstein-Yang-Mills (EYM) 
theory given by
\begin{eqnarray}
I_{EYM} = \int_{M} d^4x\sqrt{g}\left[{-1\over 16\pi}R + {1\over
4g^2_{c}}F^{a}_{\mu\nu}F^{a\mu\nu}\right] - \int_{\partial M}
d^3x\sqrt{h}{1\over 8\pi}K
\end{eqnarray}
where $F^{a}_{\mu\nu}$ is the field strength of the YM gauge field
$A^{a}_{\mu}$ with $a=1,2,3$ being the SU(2) group index and
$g_{c}$ being the gauge coupling constant. The Gibbons-Hawking
term on the boundary $\partial M$ of the manifold $M$ is also
added and $h$ is the metric induced on $\partial M$ and $K$ is the
trace of the second fundamental form on $\partial M$. Then by
extremizing this action with respect to the metric $g_{\mu\nu}$
and the YM gauge field $A^{a}_{\mu}$, one gets the following
classical field equations respectively
\begin{eqnarray}
&&R_{\mu\nu} - {1\over 2}g_{\mu\nu}R  =
8\pi T_{\mu\nu}, \nonumber \\
&&T_{\mu\nu} = {1\over
g^2_{c}} \left[F^{a}_{\mu\alpha}F_{\nu}^{a\alpha} - {1\over 4}
g_{\mu\nu}(F^{a}_{\alpha\beta}F^{a\alpha\beta})\right], \\
&&D_{\mu}\left[\sqrt{g}F^{a\mu\nu}\right] = 0, 
~~~D_{\mu}\left[\sqrt{g}\tilde{F}^{a\mu\nu}\right] = 0 \nonumber
\end{eqnarray}
where we added Bianchi identity in the last line and
$F^{a}_{\mu\nu} =
\partial_{\mu}A^{a}_{\nu}-\partial_{\nu}A^{a}_{\mu}+\epsilon^{abc}
A^{b}_{\mu}A^{c}_{\nu}$, $D^{ac}_{\mu} = \partial_{\mu}\delta^{ac}
+\epsilon^{abc}A^{b}_{\mu}$ and $A_{\mu}=A^{a}_{\mu}(-iT^{a})$,
$F_{\mu\nu}=F^{a}_{\mu\nu}(-iT^{a})$ with $T^{a}=\tau^{a}/2$
($a=1,2,3$) being the SU(2) generators and finally
$\tilde{F}_{\mu\nu} = {1\over 2}
\epsilon_{\mu\nu}^{~~\alpha\beta}F_{\alpha\beta}$ is the (Hodge) dual
of the field strength tensor. We now seek solutions ($g_{\mu\nu}$,
$A^{a}_{\mu}$) of the coupled EYM equations given above in
Euclidean signature obeying the (anti) self-dual equation in the
YM sector
\begin{eqnarray}
F^{\mu\nu} = g^{\mu\lambda}g^{\nu\sigma}F_{\lambda\sigma} = \pm
{1\over 2} \epsilon_{c}^{\mu\nu\alpha\beta}F_{\alpha\beta}
\end{eqnarray}
where
$\epsilon_{c}^{\mu\nu\alpha\beta}=\epsilon^{\mu\nu\alpha\beta}/\sqrt{g}$
is the curved spacetime version of totally antisymmetric tensor.
As was noted in [13,14,16], in Euclidean signature, the YM
energy-momentum tensor vanishes identically for YM fields
satisfying this (anti) self-duality condition. This point is of
central importance and can be illustrated briefly as follows.
Under the Hodge dual transformation, $F^{a}_{\mu\nu} \rightarrow
\tilde{F}^{a}_{\mu\nu}$, the YM energy-momentum tensor
$T_{\mu\nu}$ given in eq.(47) above is invariant normally in
Lorentzian signature. In Euclidean signature, however, its sign
flips, i.e., $\tilde{T}_{\mu\nu} = - T_{\mu\nu}$. As a result, for
YM fields satisfying the (anti) self-dual equation in Euclidean
signature such as the instanton solution, $F^{a}_{\mu\nu} = \pm
\tilde{F}^{a}_{\mu\nu}$, it follows that $T_{\mu\nu} =
-\tilde{T}_{\mu\nu} = -T_{\mu\nu}$, namely the YM energy-momentum
tensor vanishes identically, $T_{\mu\nu}=0$. This, then, indicates
that the YM field now does not disturb the geometry while the
geometry still does have effects on the YM field. Consequently the
geometry, which is left intact by the YM field, effectively serves
as a ``background'' spacetime which can be chosen somewhat at our
will (as long as it satisfies the vacuum Einstein equation
$R_{\mu\nu}=0$) and here in this work, we take it to be the
gravitational instanton. Loosely speaking, all the typical GI, including
TN metric and EH solution, possess the same
topology $R\times S^3$ and similar metric structures. Of course in a
stricter sense, their exact topologies can be distinguished, say, by different
Euler numbers and Hirzebruch signatures [21]. Particularly,
in terms of the concise basis 1-forms, the metrics of these GI can
be written as [21,22]
\begin{eqnarray}
ds^2 &=& c^2_{r}dr^2 +
c^2_{1}\left(\sigma^2_{1}+\sigma^2_{2}\right) +
c^2_{3}\sigma^2_{3} \nonumber \\ &=& c^2_{r}dr^2 +
\sum_{a=1}^{3}c^2_{a}\left(\sigma^{a}\right)^2 = e^{A}\otimes
e^{A}
\end{eqnarray}
where $c_{r}=c_{r}(r)$, $c_{a}=c_{a}(r)$, $c_{1}=c_{2}\neq c_{3}$
and the orthonormal basis 1-form $e^{A}$ is given by
\begin{eqnarray}
e^{A} = \left\{e^{0}=c_{r}dr, ~~e^{a}=c_{a}\sigma^{a}\right\}
\end{eqnarray}
and $\left\{\sigma^{a}\right\}$ ($a=1,2,3$) are the left-invariant
1-forms satisfying the SU(2) Maurer-Cartan structure equation
\begin{eqnarray}
d\sigma^{a} = -{1\over 2}\epsilon^{abc}\sigma^{b}\wedge
\sigma^{c}.
\end{eqnarray}
They form a basis on the $S^{3}$ section of the geometry and hence
can be represented in terms of 3-Euler angles $0\leq \theta
\leq\pi$, $0\leq \phi \leq 2\pi$, and $0\leq \psi \leq 4\pi$
parametrizing $S^3$ as
\begin{eqnarray}
\sigma^1 &=& -\sin\psi d\theta + \cos\psi \sin\theta d\phi, \nonumber \\
\sigma^2 &=&  \cos\psi d\theta + \sin\psi \sin\theta d\phi, \\
\sigma^3 &=& -d\psi - \cos\theta d\phi. \nonumber
\end{eqnarray}
Now in order to construct exact YM instanton solutions in the
background of these GI, we now choose the relevant ans\H{a}tz for
the YM gauge potential and the SU(2) gauge fixing. And in doing
so, our general guideline is that the YM gauge field ans\H{a}tz
should be endowed with the symmetry inherited from that of the
background geometry, the GI. Thus we first ask what kind of
isometry these GI possess. As noted above, typical GI, including
the TN and the EH metrics, possess the topology of
$R\times S^3$. The geometrical structure of the $S^3$ section,
however, is not that of perfectly ``round'' $S^3$ but rather, that
of ``squashed'' $S^3$. In order to get a closer picture of this
squashed $S^3$, we notice that the $r=$constant slices of these GI
can be viewed as U(1) fibre bundles over $S^2\sim CP^1$ with the
line element
\begin{eqnarray}
d\Omega^2_{3} = c^2_{1}\left(\sigma^2_{1}+\sigma^2_{2}\right) +
c^2_{3}\sigma^2_{3} = c^2_{1}d\Omega^2_{2} +
c^2_{3}\left(d\psi + B\right)^2
\end{eqnarray}
where $d\Omega^2_{2}=(d\theta^2 + \sin^2\theta d\phi^2)$ is the
metric on unit $S^2$, the base manifold whose volume form
$\Omega_{2}$ is given by $\Omega_{2} = dB$ as $B = \cos\theta
d\phi$ and $\psi$ then is the coordinate on the U(1)$\sim S^1$
fibre manifold. Now then the fact that $c_{1}=c_{2}\neq c_{3}$
indicates that the geometry of this fibre bundle manifold is not
that of round $S^3$ but that of squashed $S^3$ with the squashing
factor given by $(c_{3}/c_{1})$. And further, it is squashed along
the U(1) fibre direction. Thus this failure for the geometry to be
that of exactly round $S^3$ keeps us from writing down the
associated ans\H{a}tz for the YM gauge potential right away.
Apparently, if the geometry were that of round $S^3$, one would
write down the YM gauge field ans\H{a}tz as $A^{a}=f(r)\sigma^{a}$
[14,16] with $\{\sigma^{a}\}$ being the left-invariant 1-forms introduced
earlier. The rationale for this choice can be stated
briefly as follows. First, since the $r=$constant sections of the
background space have the geometry of round $S^3$ and hence
possess the SO(4)-isometry, one would look for the SO(4)-invariant
YM gauge connection ans\H{a}tz as well. Next, noticing that both
the $r=$constant sections of the frame manifold and the SU(2) YM
group manifold possess the geometry of round $S^3$, one may
naturally choose the left-invariant 1-forms $\{\sigma^{a}\}$ as
the ``common'' basis for both manifolds. Thus this YM gauge
connection ans\H{a}tz, $A^{a}=f(r)\sigma^{a}$ can be thought of as
a hedgehog-type ans\H{a}tz where the group-frame index mixing is
realized in a simple manner [14,16]. Then coming back to our present
interest, namely the GI given in eq.(49), in $r=$constant sections,
the SO(4)-isometry is partially broken down to that of SO(3) by
the squashedness along the U(1) fibre direction to a degree set by
the squashing factor $(c_{3}/c_{1})$. Thus now our task became
clearer and it is how to encode into the YM gauge connection
ans\H{a}tz this particular type of SO(4)-isometry breaking coming
from the squashed $S^3$. Interestingly, a clue to this puzzle can
be drawn from the work of Eguchi and Hanson [23] in which they
constructed abelian instanton solution in Euclidean TN
metric (namely the abelian gauge field with (anti)self-dual field
strength with respect to this metric). To get right to the point,
the working ans\H{a}tz they employed for the abelian gauge field
to yield (anti)self-dual field strength is to align the abelian
gauge connection 1-form along the squashed direction, i.e., along
the U(1) fibre direction, $A = g(r)\sigma^3$. This choice looks
quite natural indeed. After all, realizing that embedding of a
gauge field in a geometry with high degree of isometry is itself
an isometry (more precisly isotropy)-breaking action, it would be
natural to put it along the direction in which part of the
isometry is already broken. Finally therefore, putting these two
pieces of observations carefully together, now we are in the
position to suggest the relevant ans\H{a}tz for the YM gauge
connection 1-form in these GI and it is
\begin{eqnarray}
A^{a} = f(r)\sigma^{a} + g(r)\delta^{a3}\sigma^{3}
\end{eqnarray}
which obviously would need no more explanatory comments except
that in this choice of the ans\H{a}tz, it is implicitly understood
that the gauge fixing $A_{r}=0$ is taken. From this point on, the
construction of the YM instanton solutions by solving the
(anti)self-dual equation given in eq.(48) is straightforward. To
sketch briefly the computational algorithm, first we obtain the YM
field strength 2-form (in orthonormal basis) via exterior calculus
(since the YM gauge connection ans\H{a}tz is given in
left-invariant 1-forms) as $F^{a}=(F^{1}, F^{2}, F^{3})$ where
\begin{eqnarray}
F^{1} &=& {f'\over c_{r}c_{1}}(e^0\wedge e^1) + {{f[(f-1)+g]}\over
c_{2}c_{3}}(e^{2}\wedge e^{3}), \nonumber \\ 
F^{2} &=& {f'\over c_{r}c_{2}}(e^0\wedge e^2) + {{f[(f-1)+g]}\over
c_{3}c_{1}}(e^{3}\wedge e^{1}),  \\ 
F^{3} &=& {(f'+g')\over c_{r}c_{3}}(e^0\wedge e^3) + {{[f(f-1)-g]}\over
c_{1}c_{2}}(e^{1}\wedge e^{2}) \nonumber
\end{eqnarray}
from which we can read off the (anti)self-dual equation to be
\begin{eqnarray}
\pm {f'\over c_{r}c_{1}} = {{f[(f-1)+g]}\over c_{2}c_{3}},
~~~\pm {(f'+g')\over c_{r}c_{3}} = {{[f(f-1)-g]}\over c_{1}c_{2}}
\end{eqnarray}
where ``$+$'' for self-dual and ``$-$'' for anti-self-dual
equation and we have only a set of two equations as $c_{1}=c_{2}$.
The specifics of different GI are characterized by particular
choices of the orthonormal basis $e^{A} = \{e^{0}=c_{r}dr,
~~e^{a}=c_{a}\sigma^{a}\}$. Thus next, for each GI (i.e., for each
choice of $e^{A}$), we solve the (anti) self-dual equation in (56)
for ans\H{a}tz functions $f(r)$ and $g(r)$ and finally from which
the YM instanton solutions in eq.(54) and their (anti)self-dual
field strength in eq.(55) can be obtained. We now present the
solutions obtained by applying the algorithm introduced here to the 
two best-known GI, the TN and the EH metrics.

{\rm \bf (I) YM instanton in Taub-NUT metric background} 

The TN GI solution written in the metric form given in eq.(49) 
amounts to
\begin{eqnarray}
c_{r}={1\over 2}\left[{r+m\over r-m}\right]^{1/2},
~~~c_{1}=c_{2}={1\over 2}\left[r^2-m^2\right]^{1/2},
~~~c_{3}=m\left[r-m\over r+m\right]^{1/2} \nonumber
\end{eqnarray}
and it is a solution to Euclidean vacuum Einstein equation
$R_{\mu\nu}=0$ for $r\geq m$ with self-dual Riemann tensor. The
apparent singularity at $r=m$ can be removed by a coordinate
redefinition and is a `nut' (in terminology of Gibbons and Hawking
[22]) at which the isometry generated by the Killing vector
$(\partial/\partial \psi)$ has a zero-dimensional fixed point set.
And this TN instanton is an asymptotically-locally-flat (ALF) metric. \\
It turns out that only the anti-self-dual equation
$F^{a}=-\tilde{F}^{a}$ admits a non-trivial solution and it is
$A^{a}=(A^1, A^2, A^3)$ where
\begin{eqnarray}
A^1 = \pm 2{(r-m)^{1/2}\over (r+m)^{3/2}}e^1,  ~~~A^2 = \pm
2{(r-m)^{1/2}\over (r+m)^{3/2}}e^2, ~~~A^3 = {(r+3m)\over m}
{(r-m)^{1/2}\over (r+m)^{3/2}}e^3
\end{eqnarray}
and $F^{a}=(F^1, F^2, F^3)$ where
\begin{eqnarray}
F^1 &=& \pm {8m\over (r+m)^3}\left(e^0\wedge e^1 - e^2\wedge
e^3\right), ~~~F^2 = \pm {8m\over (r+m)^3}\left(e^0\wedge e^2 -
e^3\wedge e^1\right), \nonumber \\ F^3 &=&  {16m\over
(r+m)^3}\left(e^0\wedge e^3 - e^1\wedge e^2\right).
\end{eqnarray}
It is interesting to note that this YM field strength and the
Ricci tensor of the background TN GI are proportional as
$|F^{a}|=2|R^{0}_{a}|$ except for opposite self-duality, i.e.,
\begin{eqnarray}
R^0_{1}=-R^2_3 &=&  {4m\over (r+m)^3}\left(e^0\wedge e^1 +
e^2\wedge e^3\right), ~~~R^0_{2}=-R^3_1 = {4m\over
(r+m)^3}\left(e^0\wedge e^2 + e^3\wedge e^1\right), \nonumber \\
R^0_{3}=-R^1_2 &=& -{8m\over (r+m)^3}\left(e^0\wedge e^3 +
e^1\wedge e^2\right).
\end{eqnarray}

{\rm \bf (II) YM instanton in Eguchi-Hanson metric background} 

The EH GI solution amounts to
\begin{eqnarray}
c_{r}=\left[1 - \left({a\over r}\right)^4\right]^{-1/2},
~~~c_{1}=c_{2}={1\over 2}r, ~~~c_{3}={1\over 2}r\left[1 - \left({a\over
r}\right)^4 \right]^{1/2} \nonumber
\end{eqnarray}
and again it is a solution to Euclidean vacuum Einstein equation
$R_{\mu\nu}=0$ for $r\geq a$ with self-dual Riemann tensor. $r=a$
is just a coordinate singularity that can be removed by a
coordinate redefinition provided that now $\psi$ is identified
with period $2\pi$ rather than $4\pi$ and is a `bolt' (in
terminology of Gibbons and Hawking [22]) where the action of the
Killing field $(\partial/\partial \psi)$ has a two-dimensional
fixed point set. Besides, this EH instanton is an
asymptotically-locally-Euclidean (ALE) metric. \\ 
In this time, only the
self-dual equation $F^{a}=+\tilde{F}^{a}$ admits a non-trivial
solution and it is $A^{a}=(A^1, A^2, A^3)$ where
\begin{eqnarray}
A^1 = \pm {2\over r}\left[1 - \left({a\over r}\right)^4\right]^{1/2}e^1,
~~~A^2 = \pm {2\over r}\left[1 - \left({a\over r}\right)^4\right]^{1/2}e^2,
~~~A^3 = {2\over r}{\left[1 + \left({a\over r}\right)^4\right]\over \left[1 -
\left({a\over r}\right)^4\right]^{1/2}} e^3
\end{eqnarray}
and $F^{a}=(F^1, F^2, F^3)$ where
\begin{eqnarray}
F^1 &=& \pm {4\over r^2}\left({a\over r}\right)^4\left(e^0\wedge e^1 +
e^2\wedge e^3\right), ~~~F^2 = \pm {4\over r^2}\left({a\over
r}\right)^4\left(e^0\wedge e^2 + e^3\wedge e^1\right), \nonumber \\ F^3
&=& - {8\over r^2}\left({a\over r}\right)^4\left(e^0\wedge e^3 + e^1\wedge
e^2\right).
\end{eqnarray}
Again it is interesting to realize that this YM field strength and
the Ricci tensor of the background EH GI are proportional as
$|F^{a}|=2|R^{0}_{a}|$, i.e.,
\begin{eqnarray}
R^0_{1}=-R^2_3 &=& {2\over r^2}\left({a\over r}\right)^4\left(- e^0\wedge e^1
+ e^2\wedge e^3\right), ~~~R^0_{2}=-R^3_1 = {2\over r^2}\left({a\over
r}\right)^4\left(- e^0\wedge e^2 + e^3\wedge e^1\right), \nonumber \\
R^0_{3}=-R^1_2 &=& - {4\over r^2}\left({a\over r}\right)^4\left(- e^0\wedge
e^3 + e^1\wedge e^2\right).
\end{eqnarray}
The detailed anaysis of the nature of these solutions to the (anti) self-dual YM equation
in the background of TN and EH GI constructed thus far will be given in the Appendix.

\begin{center}
{\rm \bf V. Concluding remarks} 
\end{center}

In the present work, we were interested in the explicit construction of $SU(2)$ Yang-Mills instanton
solutions in the background geometry of a stack of two coincident probe $D4$-brane 
worldspaces particularly when
the metric of target spacetime in which the probe branes are embedded is given by the
Ricci-flat, magnetic extremal 4-brane solution in type IIA theory with its worldspace metric
being given by that of TN or EH gravitational instanton. This $D4$-brane
worldvolume soliton configuration was of particular interest since
with this YM instanton-gravitational instanton system on a probe
$D4$-brane worldvolume, the energy of the probe brane attains its minimum value and hence
enjoys stable state provided one employs the Tseytlin's non-abelian DBI
action for the description of multiple probe $D$-branes.
Here, for a pile of generally $N$ non-coincident Ricci-flat $D4$-branes embedded in a target
spacetime with metric given by that of TN or EH GI, it does not appear to
be totally clear whether the metric induced on the probe branes' worldspaces is that of
{\it multi-centered} TN or EH GI or just that of a single TN or EH
uniformly for each brane. If it were that of {\it multi-centered} TN or EH GI [11,24],
then the $SU(N)$ instanton solutions constructed on them should be the {\it multi}-instanton
solutions as well and would add more technical complexity to our consideration. 
Even in this case, however, if we confine our interest
to the case of a stack of $N$ {\it coincident} $D4$-branes, then things will become simpler.
That is, for a stack of $N$ {\it coincident} $D4$-branes, the centers of the $N$ GI would merge
and as a result, the metric of the $N$ coincident $D4$-branes' worldspaces would coalesce   
to become that of a single GI. Therefore, as far as the case of the stack of two {\it coincident}
Ricci-flat $D4$-branes is concerned, the corresponding worldvolume soliton configuration would be
properly described by the {\it single} $SU(2)$ YM instanton constructed on a single-centered
TN or EH GI geometry background that we discussed in the previous section.
Thus regarding the interesting observation that the BPS condition,
or the (anti) self-duality of non-abelian gauge theory instanton solution at once linearizes
the otherwise highly non-linear DBI action and minimizes the energy of the probe $D4$-branes 
(provided Tseytlin's non-abelian DBI action is employed), here in this work we have actually 
demonstrated that this procedure can actually work by constructing in an explicit manner the
$SU(2)$ YM instanton solution having (anti) self-dual field strength in the
background of {\it Ricci-flat} $D4$-branes' worldspaces with metrics given by two best-known 
4-dimensional gravitational instantons, TN and EH metrics.

\begin{center}
{\rm\bf Acknowledgments}
\end{center}

H. Kim would like to thank Prof. Sang-Jin Sin for general discussions on various aspects of 
$D$-brane physics. 
This work was supported in part by the Brain Korea 21 project and by the basic
science promotion program from Korea Research Foundation. Y. Yoon also wishes to acknowledge
financial support of Hanyang univ. made in the program year of 2000.

\vspace*{2cm}

\begin{center}
{\rm \bf Appendix : Analysis of the nature of solutions in sect.IV}
\end{center}

In this Appendix, we would like to examine the nature of the solutions to (anti) self-dual YM equation
in the background of TN and EH GI discussed in sect.IV.  First, recall
that the relevant ans\H{a}tz for the YM gauge connection is of the
form $A^{a}=f(r)\sigma^{a} + g(r)\delta^{a3}\sigma^3$ in the TN and EH GI backgrounds
with topology of $R\times ({\rm squashed})S^3$. Here, however, the physical interpretation
of the nature of YM gauge
potential solutions $A^{a}$ is rather unclear when they are expressed in terms of these
left-invariant 1-forms $\{\sigma^{a}\}$ or the orthonormal basis $e^{A}$ in eq.(50).
Thus in order to get a better insight into the
physical meaning of the structure of these YM connection
ans\H{a}tz, we now try to re-express the left-invariant 1-forms
$\{\sigma^{a}\}$ forming a basis on $S^3$ in terms of more
familiar Cartesian coordinate basis. And this can be achieved
by first relating the polar coordinates $(r, ~\theta, ~\phi, ~\psi)$ to
Cartesian $(t,x,y,z)$ coordinates (note, here, that $t$ is not
the usual ``time'' but just another spacelike coordinate) given by
[21]
\begin{eqnarray}
x+iy = r\cos {\theta\over 2}\exp{[{i\over 2}(\psi+\phi)]}, ~~~z+it =
r\sin {\theta\over 2}\exp{[{i\over 2}(\psi-\phi)]},
\end{eqnarray}
where $x^2+y^2+z^2+t^2=r^2$ which is the equation for $S^3$ with radius $r$.
From this coordinate transformation law, one now can relate the
non-coordinate basis to the Cartesian coordinate basis as
\begin{eqnarray}
\pmatrix{dr \cr
         r\sigma_{x} \cr
         r\sigma_{y} \cr
         r\sigma_{z} \cr} = {1\over r}
\pmatrix{x & y & z & t \cr
         -t & -z & y & x \cr
         z & -t & -x & y \cr
         -y & x & -t & z \cr}
\pmatrix{dx \cr
         dy \cr
         dz \cr
         dt \cr}
\end{eqnarray}
where $\{\sigma_{x}=-\sigma^{1}/2, ~\sigma_{y}=-\sigma^{2}/2,
 ~\sigma_{z}=-\sigma^{3}/2\}$.
Still, however, the meaning of YM gauge connection ans\H{a}tz rewritten
in terms of the Cartesian coordinate basis $dx^{\mu}=(dt, ~dx, ~dy, ~dz)$
as above does not look so apparent. Thus we next introduce
the so-called {\it `tHooft tensor} [6,21] defined by
\begin{eqnarray}
\eta^{a\mu\nu}=-\eta^{a\nu\mu}=(\epsilon^{0a\mu\nu}+{1\over 2}\epsilon^{abc}
\epsilon^{bc\mu\nu}).   
\end{eqnarray}
Then the left-invariant 1-forms can be cast to a more concise form
$\sigma^{a}=2\eta^{a}_{\mu\nu}(x^{\nu}/r^2)dx^{\mu}$. Therefore,
the YM instanton solution, in Cartesian coordinate basis, can be
written as
\begin{eqnarray}
A^{a} = A^{a}_{\mu}dx^{\mu} =
2\left[f(r)+g(r)\delta^{a3}\right]\eta^{a}_{\mu\nu}{x^{\nu}\over
r^2}dx^{\mu}
\end{eqnarray}
in the background of TN and EH GI with topology of $R\times ({\rm
squashed})S^3$. Now in order to appreciate the meaning of this structure,
we go back to the flat space situation. As is well-known,
the standard BPST [12] SU(2) YM instanton solution in flat space
takes the form $A^{a}_{\mu} = 2\eta^{a}_{\mu\nu}[x^{\nu}/(r^2+\lambda^2)]$
with $\lambda$ being the size of
the instanton while the meron solution which is another non-trivial
solution to the second order YM field equation found long ago by De Alfaro, Fubini,
and Furlan [20] takes the form  $A^{a}_{\mu} = \eta^{a}_{\mu\nu}(x^{\nu}/r^2)$.
Since the pure (vacuum) gauge having vanishing field strength is given by
$A^{a}_{\mu} = 2\eta^{a}_{\mu\nu}(x^{\nu}/r^2)$, the standard instanton solution
interpolates between the trivial vacuum $A^{a}_{\mu}=0$ at $r=0$ and another
vacuum represented by this pure gauge above at $r\rightarrow \infty$ and the
meron solution can be thought of as a ``half a vacuum gauge''. Unlike the instanton
solution, however, the meron solution only solves the second order YM field
equation and fails to solve the first order (anti) self-dual equation. As is
apparent from their structures given above, the meron is an unstable solution
in that it is singular at its center $r=0$ and at $r=\infty $ while the ordinary
instanton solution exhibits no singular behavior. As was pointed out originally
by De Alfaro et al. [20], in contrast to instantons whose topological charge
density is a smooth function of $x$, the topological charge density of merons
vanishes everywhere except at its center, i.e., the singular point, such that
its volume integral is half unit of topological charge $1/2$. And curiously
enough, half-integer topological charge seems to be closely related to the
confinement in the Schwinger model [25]. It is also amusing to note that a
``time slice'' through the origin, i.e., $x_{0}=0$ of the meron configuration
yields a $SU(2)$ Wu-Yang monopole [25]. Lastly, the Euclidean meron action
diverges logarithmically and perhaps needs some regularization whereas the
standard YM instanton has finite action. \\
Thus we are led to the conclusion
that the YM instanton solution in typical GI backgrounds possess the
structure of (curved space version of) meron at large (but finite) $r$.
As is well-known,
in flat spacetime meron does not solve the 1st order (anti) self-dual
equation although it does the second order YM field equation.
Thus in this sense, this result seems remarkable since it implies
that in the GI backgrounds, the (anti) self-dual YM equation admits
solutions which exhibit the configuration of meron solution at large $r$
in contrast to the flat spacetime case. And we only
conjecture that when passing from the flat ($R^4$) to GI ($R\times
S^3$) geometry, the closure of the topology of part of the
manifold appears to turn the structure of the instanton solution
from that of standard BPST into that of meron. The concrete form of
the YM instanton solutions in each of these GI backgrounds written in
terms of Cartesian coordinate basis as in eq.(66) will be given below after
we comment on one more thing to which we now turn to. \\
Namely, we would like to investigate the topological charge values of these
solutions. It has been pointed out in the
literature that both in the background of Euclidean Schwarzschild   
geometry [13] and in the Euclidean de Sitter space [14], the (anti)
instanton solutions have the Pontryagin index of $\nu[A]= \pm 1$  
and hence give the contribution to the (saddle point approximation
to) intervacua tunnelling amplitude of $\exp{[-8\pi^2/g^2_{c}]}$,
which, interestingly, are the same as their flat space
counterparts even though these curved space YM instanton solutions
do not correspond to gauge transformations of any flat space
instanton solution [12]. This unexpected and hence rather curious
property, however, turns out not to persist in YM instantons
in these GI backgrounds we studied here. In order to see this,
we begin with the careful definition of the Pontryagin index or
second Chern class in the presence of the non-trivial background
geometry of GI. \\
Consider that we would like to find an index theorem for the
manifold ($M$) with boundary ($\partial M$). Namely, we now need 
an extended version of index theorem with boundary. To this question,
an appropriate answer has been provided by Atiyah, Patodi, and
Singer (APS) [26]. According to their extended version of index theorem, 
the total index, say, of a given geometry and of a gauge field
receives contributions, in addition to that from the usual bulk
term ($V(M)$), from a local boundary term ($S(\partial M)$) and
from a non-local boundary term ($\xi (\partial M)$). The bulk term
is the usual term appearing in the ordinary index theorem without
boundary and involves the integral over $M$ of terms
quadratic in curvature tensor of the geometry and in field strength
tensor of the gauge field. The local boundary term is given by the   
integral over $\partial M$ of the Chern-Simons forms for both the
geometry and the gauge field while the non-local boundary term
is given by a constant times the ``APS $\eta$-invariant'' [21] of
the boundary. And this last non-local boundary term becomes relevant
and meaningful when Dirac spinor field is present and interacts with
the geometry and the gauge field. Now specializing to the case at hand in
which we are interested in the evaluation of the instanton
number or the second Chern class of the YM gauge field {\it alone},
we only need to pick up the terms in the gauge sector in this APS
index theorem which reads [21]
\begin{eqnarray}
\nu[A] = Ch_{2}(F) = {-1\over 8\pi^2}[\int_{M=R\times S^3}tr(F\wedge F) -
\int_{\partial M=S^3}tr(\alpha \wedge F)|_{r=r_{0}}]
\end{eqnarray}
where $\alpha \equiv (A-A')$ is the ``second fundamental form'' {\it at} the
boundary $r=r_{0}$ and by definition [21] $A'$ has only {\it tangential}
components on the boundary $\partial M=S^3$. Recall, however, that our
choice of ans\H{a}tz for the YM gauge connection involves the gauge fixing
$A_{r}=0$ as we mentioned earlier. Namely, both $A$ and $A'$ possess only
tangential components (with respect to the $r=r_{0}$ boundary) at any
$r=r_{0}$ and hence $\alpha \equiv (A-A') = 0$ identically there. As a result,
even in the presence of the boundaries, the terms in the YM gauge sector
in the APS index theorem remain the same as in the case of index theorem with
no boundary, namely, only the bulk term survives in eq.(67) above. Thus what remains
is just a straightforward computation of this bulk term and it becomes easier
when performed in terms of orthonormal basis
$e^{A} = \left\{e^{0}=c_{r}dr, ~~e^{a}=c_{a}\sigma^{a}\right\}$, in which case,
\begin{eqnarray}
tr(F\wedge F) &=& {1\over 2}(F^{a}\wedge F^{a}) = {1\over 2}({1\over 4})
\epsilon_{ABCD}F^{a}_{AB}F^{a}_{CD} \sqrt{g}d^4x \nonumber \\
&=& (F^{1}_{01}F^{1}_{23}+F^{2}_{02}F^{2}_{31}+F^{3}_{03}F^{3}_{12})\sqrt{g}d^4x, \\
\int_{M=R\times S^3}d^4x\sqrt{g} &=& \int_{R}dr (c_{r}c_{1}c_{2}c_{3})\int^{4\pi}_{0}
d\psi \int^{2\pi}_{0}d\phi \int^{\pi}_{0}d\theta \sin \theta \nonumber \\
&=& 16\pi^{2} \int_{R} dr (c_{r}c_{1}c_{2}c_{3}) \nonumber
\end{eqnarray}
where we used $\sqrt{g}=|det e|=c_{r}c_{1}c_{2}c_{3}\sin \theta$. The period for the
$U(1)$ fibre coordinate $\psi$ for the EH metric, however, is $2\pi$ rather than
$4\pi$ to remove the bolt singularity at $r=a$ as we mentioned earlier. This completes
the description of the method for computing the topological charge of each solution.

{\rm \bf (1) YM instanton in Taub-NUT metric background} \\
In terms of the ans\H{a}tz functions $f(r)$ and $g(r)$ for the YM gauge connection   
in GI backgrounds given in eq.(54), the standard instanton solutions in TN metric
amount to
\begin{eqnarray}
f(r) &=& \left({r-m \over r+m}\right), ~~~g(r) = \left({2m\over r+m}\right)
\left({r-m \over r+m}\right), \\
f(r) &=& -\left({r-m \over r+m}\right), ~~~g(r) = 2\left({r+2m\over r+m}\right)   
\left({r-m \over r+m}\right) \nonumber
\end{eqnarray}  
for self-dual and anti-self-dual YM equations respectively.
Therefore, when expressed in Cartesian coordinate basis as in eq.(66),
the solutions take the forms
\begin{eqnarray}
A^{a}_{\mu} &=& 2\left({r-m \over r+m}\right)\left[1 + \left({2m\over r+m}\right)
\delta^{a3}\right]\eta^{a}_{\mu\nu}{x^{\nu}\over r^2}, \\
A^{a}_{\mu} &=& 2\left({r-m \over r+m}\right)\left[-1 + 2\left({r+2m\over r+m}\right)
\delta^{a3}\right]\eta^{a}_{\mu\nu}{x^{\nu}\over r^2} \nonumber
\end{eqnarray}
for self-dual and anti-self-dual case respectively. Some comments regarding the
features of these solutions are now in order. i) They appear to be singular at the
center $r=0$ but it should not be a problem as $r\geq m$ for the background TN metric
and hence the point $r=0$ is absent. ii) It is interesting to note that the solutions
become vacuum gauge $A^{a}_{\mu}=0$ at the boundary $r=m$ which has the topology of
$S^3$. iii) For $r\rightarrow \infty$, the solutions asymptote to another vacuum gauge
$|A^{a}_{\mu}| = 2\eta^{a}_{\mu\nu}(x^{\nu}/r^2)$. \\   
We now turn to the computation of the topological charge, i.e., the Pontryagin index
of these YM solution. The relevant quantities involved in this computation are the
ones in eq.(68) and they, for the case at hand, are
\begin{eqnarray}
(c_{r}c_{1}c_{2}c_{3}) &=& {m\over 8}(r^2-m^2), \\
F^{a}_{\mu\nu}\tilde{F}^{a \mu\nu} &=&
4(F^{1}_{01}F^{1}_{23}+F^{2}_{02}F^{2}_{31}+F^{3}_{03}F^{3}_{12})
= -24{(8m)^2\over (r+m)^6}. \nonumber
\end{eqnarray}  
Thus we have
\begin{eqnarray}
\nu[A] &=& \left({-1\over 32\pi^2}\right)16\pi^2 \int^{\infty}_{m}dr {m\over 8}(r^2-m^2)
\left[-24{(8m)^2\over (r+m)^6}\right] \nonumber \\
&=& 1.
\end{eqnarray}

{\rm \bf (2) YM instanton in Eguchi-Hanson metric background} \\
The standard instanton solutions in EH metric amount to
\begin{eqnarray}
f(r) &=& \left[1-\left({a\over r}\right)^4\right]^{1/2},
~~~g(r) = \left[1+\left({a\over r}\right)^4\right] -
\left[1-\left({a\over r}\right)^4\right]^{1/2}, \\ 
f(r) &=& -\left[1-\left({a\over r}\right)^4\right]^{1/2},
~~~g(r) = \left[1+\left({a\over r}\right)^4\right] +
\left[1-\left({a\over r}\right)^4\right]^{1/2}  \nonumber
\end{eqnarray}
for self-dual and anti-self-dual YM equations respectively.
Thus in Cartesian coordinate basis, the solutions take the forms
\begin{eqnarray}
A^{a}_{\mu} &=& 2\left\{\left[1-\left({a\over r}\right)^4\right]^{1/2} +
\left(\left[1+\left({a\over r}\right)^4\right] -
\left[1-\left({a\over r}\right)^4\right]^{1/2}\right)\delta^{a3}\right\}
\eta^{a}_{\mu\nu}{x^{\nu}\over r^2}, \\
A^{a}_{\mu} &=& 2\left\{-\left[1-\left({a\over r}\right)^4\right]^{1/2} +
\left(\left[1+\left({a\over r}\right)^4\right] +
\left[1-\left({a\over r}\right)^4\right]^{1/2}\right)\delta^{a3}\right\}
\eta^{a}_{\mu\nu}{x^{\nu}\over r^2} \nonumber
\end{eqnarray}  
for self-dual and anti-self-dual cases respectively. Some comments regarding the
features of these solutions are now in order. i) Again, they appear to be singular at the
center $r=0$ but it should not be a problem as $r\geq a$ for the background EH metric
and hence the point $r=0$ is absent. ii) The solutions become
$A^{a}_{\mu}=4\eta^{a}_{\mu\nu}\delta^{a3}(x^{\nu}/r^2)$ at the boundary $r=a$ which
has the topology of $S^3/Z_{2}$.
iii) For $r\rightarrow \infty$, the solutions asymptote to the vacuum gauge
$|A^{a}_{\mu}| = 2\eta^{a}_{\mu\nu}(x^{\nu}/r^2)$. \\
We turn now to the computation of the Pontryagin index of these YM solution.
For the case at hand, the relevant quantities involved in this computation are
\begin{eqnarray}
(c_{r}c_{1}c_{2}c_{3}) = {1\over 8}r^3,
~~~F^{a}_{\mu\nu}\tilde{F}^{a \mu\nu} = 24\left({4a^4\over r^6}\right)^2.
\end{eqnarray}
Thus we have
\begin{eqnarray}
\nu[A] = \left({-1\over 32\pi^2}\right)8\pi^2 \int^{\infty}_{a}dr {1\over 8}r^3
\left[24\left({4a^4\over r^6}\right)^2\right]
= -{3\over 2}   
\end{eqnarray}
where we set the range for the $U(1)$ fibre coordinate as $0\leq \psi \leq 2\pi$
rather than $0\leq \psi \leq 4\pi$ for the reason stated earlier.
Note particularly that it is precisely this point that renders the Pontryagin index
of this solution {\it fractional} because otherwise, it would come out as $-3$ instead. \\
Let us now discuss the behavior of these solutions as $r\rightarrow 0$ once again
to stress that they really do not exhibit singular behaviors there.
For TN and EH instantons, the
ranges for radial coordinates are $m\leq r <\infty$ and $a\leq r<\infty$, respectively.
Since the point $r=0$ is absent in these
manifolds, the solutions in these GI are everywhere regular.
At large but finite $r$, on the other hand, the solutions appear to take the structure
close to that of meron solution in flat space. Another interesting point worthy of note is
that the solution in TN background exhibits a generic
property of the instanton solution in that it does interpolate between a trivial vacuum
at $r=m$ and another vacuum (pure gauge) at $r\rightarrow \infty $.
Namely, the solution in TN background appears to exhibit
features of both meron such as their large $r$ behavior
and instanton such as interpolating configuration between two vacua.
Next, we analyze the meaning of the topological charge values of
the solutions. It is remarkable that generally the solutions seem to carry
{\it fractional} topological charge values.
Here, however, the solution in EH metric background carries the half-integer Pontryagin
index actually because the range for the $U(1)$ fibre coordinate is
$0\leq \psi \leq 2\pi$ and hence the boundary of EH space is $S^3/Z_2$.
To summarize, the solution in TN background particularly displays features generic in the
standard instanton while in the case of that in EH background, such
generic features of the instanton is somewhat obscured by meron-type natures.
There, however, is one obvious consensus. Both solutions in these GI
backgrounds are non-singular at their centers and have finite Euclidean
YM action. And this last point allows us to suspect that these solutions
are more like instantons in their generic nature although looks rather like
merons in their structures.

\noindent

\begin{center}
{\rm\bf References}
\end{center}

\begin{description}

\item {[1]} J. Polchinski, {\it TASI Lectures on $D$-branes}, {\it hep-th/9611050}.
\item {[2]} See for instance, K. G. Savvidy, {\it Born-Infeld Action in String Theory},
            {\it hep-th/9906075} and references therein.
\item {[3]} E. Witten, Nucl. Phys. {\bf B460}, 335 (1996).
\item {[4]} A. A. Tseytlin, Nucl. Phys. {\bf B501}, 41 (1997).
\item {[5]} P. C. Argyres and C. R. Nappi,  Nucl. Phys. {\bf B330}, 151 (1990).
\item {[6]} J. P. Gauntlett, J. Gomis, and P. K. Townsend, JHEP {\bf 01}, 003 (1998) ;
            G. W. Gibbons and K. Hashimoto, JHEP {\bf 09}, 013 (2000).
\item {[7]} G. W. Gibbons, Nucl. Phys. {\bf B514}, 603 (1998).
\item {[8]} C. G. Callan and J. Maldacena,  Nucl. Phys. {\bf B513}, 198 (1998).
\item {[9]} K. Dasgupta and S. Mukhi,  Phys. Lett. {\bf B423}, 261 (1998).
\item {[10]} D. Brecher,  Phys. Lett. {\bf B442}, 117 (1998).
\item {[11]} D. Brecher and M. J. Perry,  Nucl. Phys. {\bf B566}, 151 (2000).
\item {[12]} A. A. Belavin, A. M. Polyakov, A. S. Schwarz, and Yu. S. Tyupkin,
            Phys. Lett. {\bf B59}, 85 (1975) ;
            G. `tHooft, Phys. Rev. Lett. {\bf 37}, 8 (1976).
\item {[13]} J. M. Charap and M. J. Duff, Phys. Lett. {\bf B69}, 445 (1977) ;
            {\it ibid} {\bf B71}, 219 (1977).
\item {[14]} H. Kim and S. K. Kim,
            Nuovo Cim. {\bf B114}, 207 (1999) and references therein.
\item {[15]} M. Atiyah, V. Drinfeld, N. Hitchin, and Y. Manin, Phys. Lett. {\bf A65},
            185 (1987) ; P. B. Kronheimer and H. Nakajima, Math. Ann. {\bf 288}, 263 (1990).
\item {[16]} H. Kim and Y. Yoon, Phys. Lett. {\bf B495}, 169 (2000) ({\it hep-th/0002151}) ;
             H. Kim and Y. Yoon, {\it hep-th/0012055}.
\item {[17]} K. S. Stelle, ICTP Lectures, {\it hep-th/9803116}.
\item {[18]} A. Taub, Ann. Math. {\bf 53}, 472 (1951) ;
            E. Newman, L. Tamburino, and T. Unti, J. Math. Phys. {\bf 4}, 915 (1963) ;
            S. W. Hawking, Phys. Lett. {\bf A60}, 81 (1977).
\item {[19]} T. Eguchi and A. J.Hanson, Phys. Lett. {\bf B74}, 249 (1978).
\item {[20]} V. De Alfaro, S. Fubini, and G. Furlan, Phys. Lett. {\bf B65}, 163 (1976).
\item {[21]} T. Eguchi, P. B. Gilkey, and A. J. Hanson, Phys. Rep. {\bf 66}, 213 (1980).
\item {[22]} G. W. Gibbons and C. N. Pope, Commun. Math. Phys. {\bf 66}, 267 (1979) ;
            G. W. Gibbons and S. W. Hawking, {\it ibid}, {\bf 66}, 291 (1979).
\item {[23]} T. Eguchi and A. J. Hanson, Ann Phys. {\bf 120}, 82 (1979).
\item {[24]} S. W. Hawking, Phys. Lett. {\bf A60}, 81 (1977) ; 
             G. W. Gibbons and  S. W. Hawking, Phys. Lett. {\bf B78}, 430 (1978).
\item {[25]} C. G. Callan, R. Dashen, and D. J. Gross, Phys. Rev. {\bf D17}, 2717 (1978).
\item {[26]} M. F. Atiyah, V. K. Patodi, and I. M. Singer, Bull. London Math. Soc. {\bf 5},
             229 (1973) ; Proc. Camb. Philos. Soc. {\bf 77}, 43 (1975) ; {\it ibid.} {\bf 78},
             405 (1975) ; {\it ibid.} {\bf 79}, 71 (1976).

\end{description}

\end{document}